\documentclass[twocolumn,superscriptaddress,showpacs,amsmath,floatfix,notitlepage]{revtex4-1}
\usepackage{amsmath}
\usepackage{amssymb}
\usepackage{bm}
\usepackage{epsfig}
\usepackage{graphicx}
\usepackage{color}
\usepackage[colorlinks]{hyperref}
\usepackage[T2A]{fontenc}
\usepackage[cp1251]{inputenc}
\usepackage[english]{babel}

\usepackage{array}
\newcolumntype{x}[1]{>{\centering\let\newline\\\arraybackslash\hspace{0pt}}p{#1}}
\usepackage{multirow}

\usepackage[bbgreekl]{mathbbol}

\newcount\timehh  \newcount\timemm
\timehh=\time \divide\timehh by 60
\timemm=\time
\begin{document}

\title{Scaling laws of Rydberg excitons}

\author{J. Heck\"otter}
\affiliation{Experimentelle Physik 2, Technische Universit\"{a}t
Dortmund, D-44221 Dortmund, Germany}
\author{M. Freitag}
\affiliation{Experimentelle Physik 2, Technische Universit\"{a}t
Dortmund, D-44221 Dortmund, Germany}
\author{D. Fr\"ohlich}
\affiliation{Experimentelle Physik 2, Technische Universit\"{a}t
Dortmund, D-44221 Dortmund, Germany}
\author{M. A\ss mann}
\affiliation{Experimentelle Physik 2, Technische Universit\"{a}t
Dortmund, D-44221 Dortmund, Germany}
\author{M. Bayer}
\affiliation{Experimentelle Physik 2, Technische Universit\"{a}t
Dortmund, D-44221 Dortmund, Germany}
\affiliation{Ioffe Institute,
Russian Academy of Sciences, 194021, St.-Petersburg, Russia}

\author{M. A. Semina}
\affiliation{Ioffe Institute, Russian Academy of Sciences, 194021,
St.-Petersburg, Russia}
\author{M. M. Glazov}
\affiliation{Ioffe Institute, Russian Academy of Sciences, 194021,
St.-Petersburg, Russia}

\begin{abstract}
Rydberg atoms have attracted considerable interest due to their huge interaction among each other and with external fields. They demonstrate characteristic scaling laws in dependence on the principal quantum number $n$ for features such as the magnetic field for level crossing or the electric field of dissociation. Recently, the observation of excitons in highly excited states has allowed studying Rydberg physics in cuprous oxide crystals. While bearing striking similarities to Rydberg atoms, also fundamentally new insights may be obtained for Rydberg excitons, as the crystal environment and the associated symmetry reduction compared to vacuum give easy optical access to many states within an exciton multiplet. Here we study experimentally and theoretically the scaling  of several characteristic parameters of Rydberg excitons with $n$ providing also insight into interaction effects. From absorption spectra in magnetic field we find for the first crossing of levels with adjacent principal quantum numbers a $B_r \propto n^{-4}$ dependence of the resonance field strength, $B_r$, due to the dominant paramagnetic term unlike in the atomic case where the diamagnetic contribution is decisive resulting in a $B_r \propto n^{-6}$ dependence. By contrast, in electric field we find scaling laws just like for Rydberg atoms. The resonance electric field strength shows a scaling as $E_r \propto n^{-5}$. Similar to Rydberg atoms with the exception of hydrogen we observe anticrossings of the states belonging to multiplets with different principal quantum numbers. The energy splittings at the avoided crossings scale as $n^{-4}$ which we relate to the crystal specific deviation of the exciton Hamiltonian from the hydrogen model. The data also allow us to assess the susceptibility of Rydberg exciton to the external fields: We observe the exciton polarizability in the electric field to scale as $n^7$. In magnetic field, on the other hand, the crossover field strength from a hydrogen-like exciton to a magnetoexciton dominated by electron and hole Landau level quantization scales as $n^{-3}$. Also the exciton ionization can be studied with ionization voltages that demonstrate a $n^{-4}$ scaling law as for atoms. Particularly interesting is the field dependence of the width of the absorption lines which remains constant before dissociation for high enough $n$, while for small $n \lesssim 12$ an exponential increase is found. These results are in excellent agreement with theoretical calculations.
\end{abstract}

\maketitle

\section{Introduction}

The recent observation of highly excited excitons in bulk cuprous oxide crystals up to principal quantum numbers of $n = 25$ has opened appealing perspectives for studying Rydberg physics in the solid state~\cite{Kazimierczuk:2014yq}. The exciton Rydberg energy, $\mathcal R$, of about $90$~meV is reduced by more than two orders of magnitude as compared to the atomic Rydberg energy. This reduction makes excitons quite susceptible to external fields and enables reaching the strong field regime where the characteristic interaction energies with an electric or magnetic field prevail over the Coulomb interaction. Furthermore, the different states of the hydrogen-like excitonic series are spaced rather closely facilitating systematic high resolution studies because the absorption across the whole series can be recorded by tuning a single light source such as a frequency stabilized laser with neV-resolution.

These opportunities have been exploited already to elaborate similarities and differences of Rydberg systems of atoms and excitons. A particular example is the quantum defect, which for atoms describes the deviation of the energy levels of the valence electron from a pure hydrogen series. Similar deviations were observed for Rydberg excitons in cuprous oxide whose binding energies can be also approximately cast into the form~\cite{PhysRevB.93.075203}:
\begin{equation}
\label{qdef}
E_{n,l} = \frac{{\mathcal R}}{n^2 \left(1-\frac{\delta_{n,l}}{n} \right)^2},
\end{equation}
where $\delta_{n,l}$ is the quantum defect of the exciton in the state with principal quantum number $n$ and orbital angular momentum $l$ of the envelope wave function. Despite of this similarity the origin of the quantum defect is very different: For atoms it accounts for the shielding of the Coulomb potential of the nucleus by the closed shells of inner electrons, leading to an effective deviation from the $1/r$-dependence. In semiconductors like Cu$_2$O, it results predominantly from the details of the valence band structure, which has a dispersion deviating strongly from a parabolic law.

This exemplary finding asks for a more detailed study of the two different classes of Rydberg objects. Particularly useful in this respect is the application of external electric and magnetic fields, which cause characteristic shifts of the energy levels and also lift level degeneracies. For atoms, e.g., there is still a degeneracy in the magnetic quantum number, $m$, caused by the spherical symmetry of the system. This degeneracy can be lifted by a magnetic field. Unlike for atoms, the anisotropic crystal structure makes this description approximative only for excitons as demonstrated by the observation of a triplet splitting of the $F$-excitons, $l=3$, in absence of an external field~\cite{PhysRevLett.115.027402}.

While the effects of external fields were investigated already right after the first observation of excitons in cuprous oxide~\cite{0038-5670-5-2-A03,PSSB:PSSB2220660140,PSSA:PSSA2210430102}, recently corresponding studies were performed in both electric~\cite{efield} and magnetic~\cite{bfield} field with unprecedented resolution. Elaborated calculations accounting for the details of the band structure allowed a quantitative description of the observed field dispersions in the regime of low $n \leq 6$, where the exciton lines can be reliably assigned. For higher $n$ the field induced splittings become so diverse that statistical methods have to be applied for the level analysis. In magnetic field, this analysis revealed quantum chaotic behavior~\cite{Aszmann:2016aa}.

The focus here is placed on the high-$n$ Rydberg regime ($n > 6$) addressing quantities that can be uniquely determined and are therefore characteristic for the exciton states, in particular in an external electric or magnetic field. These quantities are the field strengths of the first resonance of states belonging to adjacent manifolds with principal quantum numbers $n$ and $n+1$, and in electric field the polarizability, the spectral linewidth, and the dissociation field strength.  A systematic picture is developed by scanning the excited state resonances from low $n$ up to the band gap, from which similarities to atoms can be extracted such as the $n^{-5}$ scaling of the first resonance on voltage in electric field~\cite{gallagher2005rydberg}. At these resonances anticrossings are observed which we attribute mainly to non-parabolic terms in the kinetic energy of the valence band hole, which represents a crystal specific feature. At higher field strengths, exciton dissociation takes place at voltages that scale as $n^{-4}$ like in the atomic case. On the other hand, in magnetic field we find a scaling different from the $n^{-6}$-law in atoms, namely a $n^{-4}$  law of the resonance field strength. This difference arises from the modification of the selection rules for excitons as compared with atoms which basically makes the whole exciton multiplet accessible in optical absorption. As a result, the states with large paramagnetic energy shift are involved in the resonances in contrast to the dominant diamagnetic behavior for atoms. We also address the multiplet width at zero field as a function of $n$.

After presenting the experimental details in the next Section~\ref{sec:exper}, we discuss the scaling law behavior in Section~\ref{sec:scaling}, considering first the zero-field situation and then turning to the behavior in external fields. In each part we also give the corresponding theoretical analysis. The manuscript is concluded in Sec. \ref{sec:concl}.

\section{Experiment}\label{sec:exper}

The experiments were performed on bulk-like cuprous oxide crystal slabs that were cut and polished from a natural rock.  
The samples were placed strain-free in a sample holder that allowed application of an electric field \cite{Brandt07}. The holder was placed at $T = 1.3$~K in the liquid helium insert of an optical cryostat equipped with a magnet coil for fields up to 7 T. Both electric and magnetic field were applied in longitudinal field configuration, i.e. they were oriented along the optical axis normal to the crystal slab plane. Further technical details can be found in Refs.~\cite{efield,bfield}.

The absorption was measured, depending on the required spectral resolution, by using either a broadband white light source or a frequency stabilized dye laser with neV-linewidth for excitation. For the white light source the bandwidth was reduced to the wavelength range of interest by a double monochromator. 
After transmission through the crystal the white light was dispersed by another double monochromator and detected by a Si-charge coupled device camera. The spectral resolution provided is less than 10~$\mu$eV which is sufficient for exciton states with $n < 20$ to study the effects of interest here. 
When the laser was used to obtain a particularly high spectral resolution, its output was stabilized by a noise eater. After transmission through the sample it was detected by a avalanche photodiode. The power density of the exciting light was chosen low enough that the excitation of dressed states as discussed in Ref.~\cite{PhysRevLett.117.133003} can be neglected and the experimentally observed spectral lines correspond to resonant absorption peaks.

Recently we showed that the absorption spectra of cuprous oxide crystals vary with the light polarization in combination with the chosen crystal orientation. For example, in electric field applied to a $[110]$ oriented crystal, the quadrupole allowed transitions to the $S$- and $D$-excitons which can be observed for light polarized linearly along the $[001]$-direction are absent for light polarized along the $[1\bar10]$ direction, simplifying the spectra significantly~\cite{efield}.  Depending on the problem to be studied, we chose the appropriate polarization configuration. If the task was to address the whole multiplet belonging to a particular $n$ for which as many states as possible need to be resolved in a manifold, we used the $[001]$-polarization, while the $[1\bar10]$-polarization was chosen when, for example, the field dispersion of the $P$-excitons was to be measured with high accuracy.

For the subsequent discussion several remarks are due. For atoms, in a constant external field the spherical symmetry is reduced, but the rotational symmetry about the field is maintained, so that the angular momentum component along the field direction, $m$, remains a good quantum number. In a longitudinal field configuration, where the optical axis is oriented along the field, well defined selection rules hold in electric dipole approximation for single photon absorption: Starting from an electron in a $s$-shell, only transitions to $p$-shell states are possible due to the restriction $\Delta l = \pm 1$, whereby all states in this state triplet can be excited leaving the magnetic quantum number unchanged, $\Delta m=0$, or changing it by unity, $\Delta m = \pm 1$, for linearly or circularly polarized light, respectively. Other states of the Rydberg manifold with higher angular momenta $l > 1$, for example, cannot be excited out of a $s$-shell state.

The situation is different for the yellow excitons in cuprous oxide, where the dominant dipole-allowed excitons are those with $P$-type envelope wavefunction. The absorption of light takes place through excitation of an exciton in a particular quantum state out of the ground state of the crystal. Due to the reduction from continuous to discrete symmetry oscillator strength is shuffled from the $P$-states to other odd exciton states, namely the $F$-, $H$-, $\ldots$ excitons, even though it remains at least two orders of magnitude smaller compared to the $P$-states. Furthermore, the $S$-exciton transitions are allowed in quadrupole approximation, and, through the exchange coupling, also the $D$-excitons become accessible. Thereby, if allowed by the transition matrix elements, the observation of states is facilitated or enhanced at the expense of the $P$-excitons, as described quantitatively in recent studies of low lying excitons~\cite{efield,bfield}.

\begin{figure*}[ht]
\includegraphics[width=\textwidth]{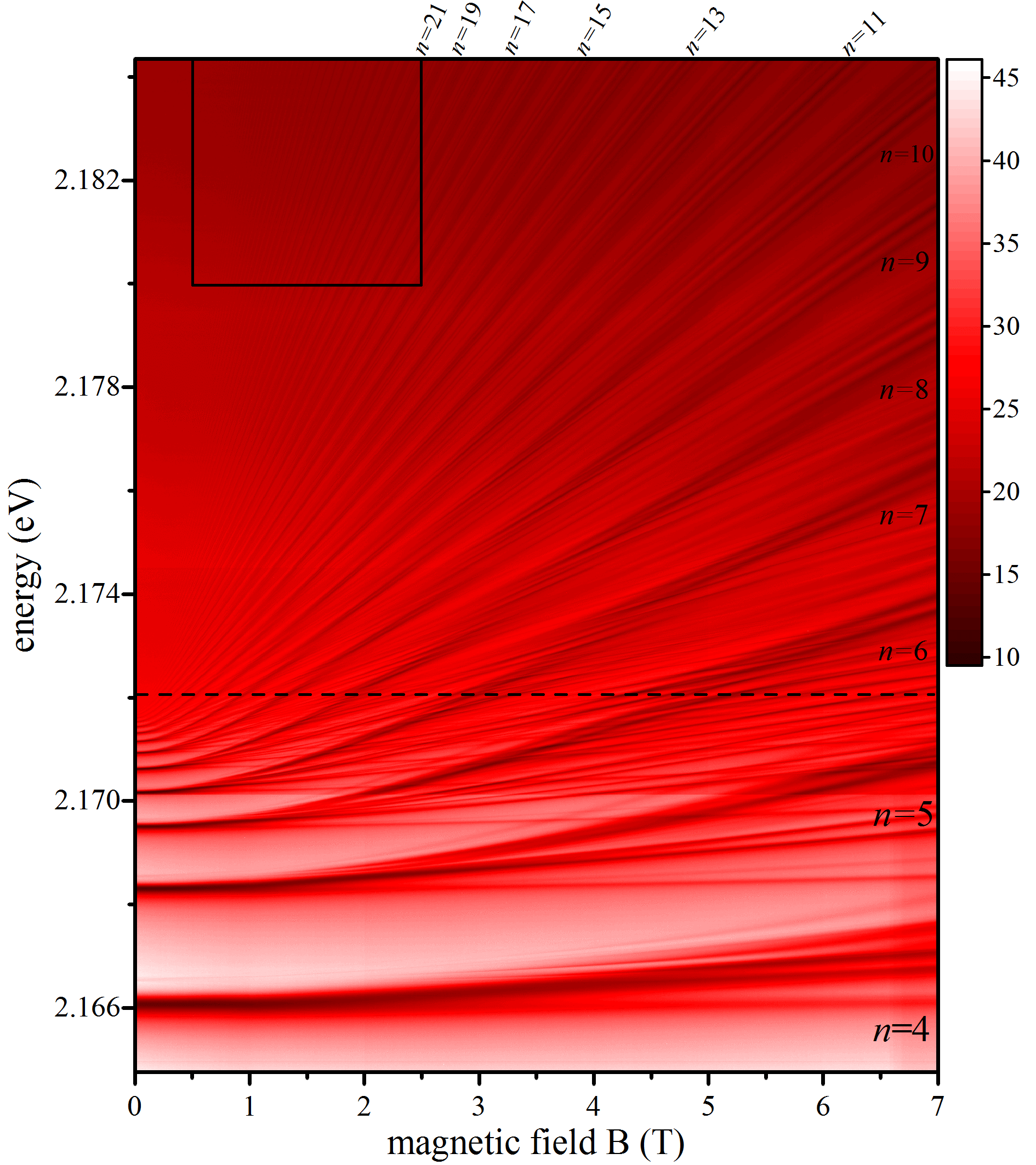}
\caption{Contour plot of absorption spectra versus magnetic field applied in the Faraday configuration (light propagation along the magnetic field), recorded at $T$ = 1.3~K on a Cu$_2$O crystal slab. The dashed line indicates the band gap at zero field. Landau level quantum numbers that can be approximately assigned to a bunch of clustered transitions are indicated at the right and top axes. The black frame indicates the area from which a close-up is shown in the Appendix \ref{app:ll} as Fig.~\ref{Fig-Appendix-LandauLevels}. The scale on the right shows gives the strength of the absorption features in arbitrary units.}\label{Fig1-bfield-spectra}
\end{figure*}

In more detail, a magnetic field enhances the visibility of the weak states that are accessible through the crystal symmetry-induced mixing of excitons with odd parity envelope as this mixing is strongly increased by the field. Quadrupole-allowed excitons with even parity  
remain weak~\cite{bfield}. By contrast, in electric field the excitons with opposite parity become mixed from which the even parity excitons profit in oscillator strength. Hence, the $S$-, $D$-, \ldots excitons become more pronounced in the absorption spectra. As a consequence of this mixing by the crystal structure and by the external field, the majority of states within an exciton multiplet becomes easily detectable, in contrast to atoms. This allows for access also to high angular momentum states by single photon absorption. 

\section{Scaling of exciton properties}\label{sec:scaling}

\begin{figure*}[ht]
\includegraphics[width=1\textwidth]{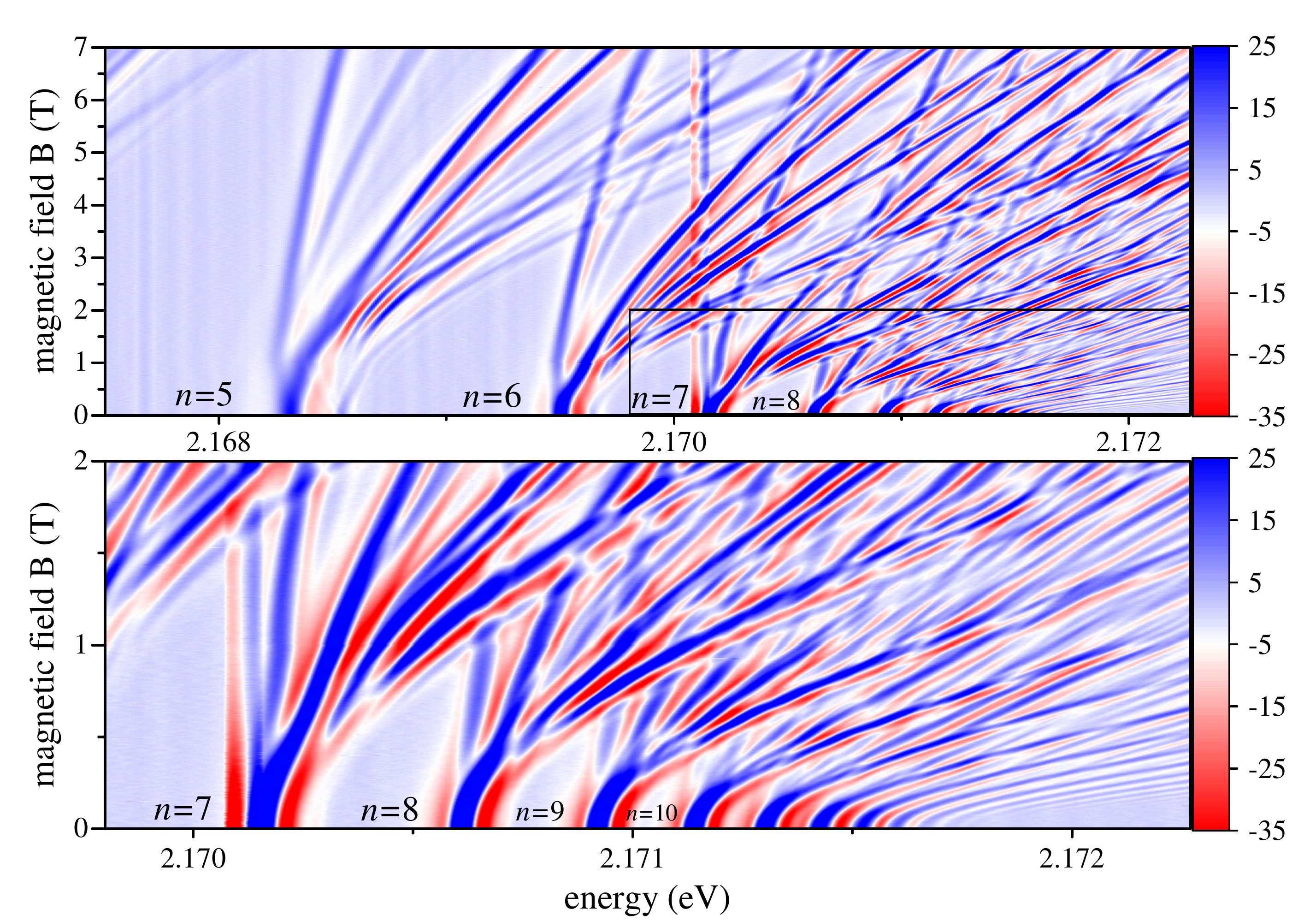}
\caption{Top panel: Same data as in Fig.~\ref{Fig1-bfield-spectra}, but in form of a contour plot of the second derivative of the absorption spectra versus magnetic field for $n \geq 5$. Lower panel: Close-up of the states for $n \geq 7$ in $B$ up to 2~T. $T$ = 1.3~K. The scales on the right give the strength of the features  in arbitrary units. The weak equidistant vertical stripes which are apparent mostly at low energies in the upper panel, are artifacts of taking the 2nd derivative.}\label{Fig2-bfield-spectra-2ndderivative}
\end{figure*}

Before discussing the different scaling laws, we first present briefly the comprehensive set of data from which the laws are extracted. These data comprise spectra recorded in magnetic and electric field, from which also the understanding of the zero field level spectrum can be deepened. Figure~\ref{Fig1-bfield-spectra} shows a contour plot of exciton absorption spectra versus the magnetic field applied in the Faraday geometry. At zero field the hydrogen-like series of exciton states is seen, out of which a complex Zeeman splitting pattern arises. Better insight into the details of these spectra can be obtained by taking their second derivatives from which weak features become well resolved. The corresponding contour plot of the second derivatives is shown in Fig.~\ref{Fig2-bfield-spectra-2ndderivative}: in panel (a) for the states with $n>4$ up to $B$ = 7~T, while in panel (b) a close-up of the low-field range up to 2~T is shown for $n>6$. At zero field the levels belonging to a particular principal quantum number are spread over an energy range which becomes the narrower, the higher $n$ is. However, in magnetic field the splitting of a multiplet becomes larger for higher $n$, because higher angular momentum states limited by $l \leq n-1$ contribute. 

Due to the multitude of observed levels in particular at medium field strengths it appears difficult to extract generally valid scaling laws for the dependences on the principal quantum number there. For that purpose we have to restrict our analysis to ranges, where states can be identified well. This is the case 
at low fields up to the point, where resonances of states belonging to exciton multiplets with different principal quantum numbers occur.  
Similarly, states can be quite well identified in high magnetic fields, where the dominant observed lines tend to cluster around transitions that correspond to those between Landau levels, see Fig.~\ref{Fig1-bfield-spectra}. In fact, the close-up of Fig.~\ref{Fig1-bfield-spectra} shown in the Appendix~\ref{app:ll} as Fig.~\ref{Fig-Appendix-LandauLevels} reveals transitions which can be associated with Landau level quantum numbers up to 79, which arise in magnetic field from $P$-excitons with $n$=79 in zero field, see also discussion below in Sec.~\ref{sec:magn}. The corresponding average radius of this exciton wavefunction would be 10.4~$\mu$m, which is squeezed in magnetic field, thereby enhancing the oscillator strength so that these highly excited states become visible. The $n$=79 state can be observed starting from 0.5~T, where the Landau level extension is about 300~nm.

Exciton level splitting can be also induced by applying an electric field and exploiting the Stark effect. Corresponding spectra are shown in Fig.~\ref{Fig3-efield-spectra} where their second derivatives are plotted as function of the voltage applied to the sample for the excitons with $n \geq 5$. Overall, the number of lines observed there remains smaller than in magnetic field because already for pretty low voltages below 10~V the exciton states with $n >$10 are subject to field-induced dissociation. Still, for identifying scaling laws, one has to restrict also here to the low field strength regime or to lines with  dominant oscillator strength compared to the other features. As discussed above, variation of the linear polarization of the exciting light allows one to vary the number of detected spectral lines. In the left panel, the light polarization was $\hat{\bm e} \parallel [1\bar{1}0]$ and in the right panel the polarization was $\hat{\bm e}\parallel [001]$. In the former case, the quadrupolar-active, even exciton states (e.g., $S$- and $D$-excitons) are forbidden, while in the latter case they are allowed, giving the spectra a more complex appearance.

\subsection{Zero-field behavior}\label{sec:zero}

\begin{figure*}[ht]
\includegraphics[width=1\linewidth]{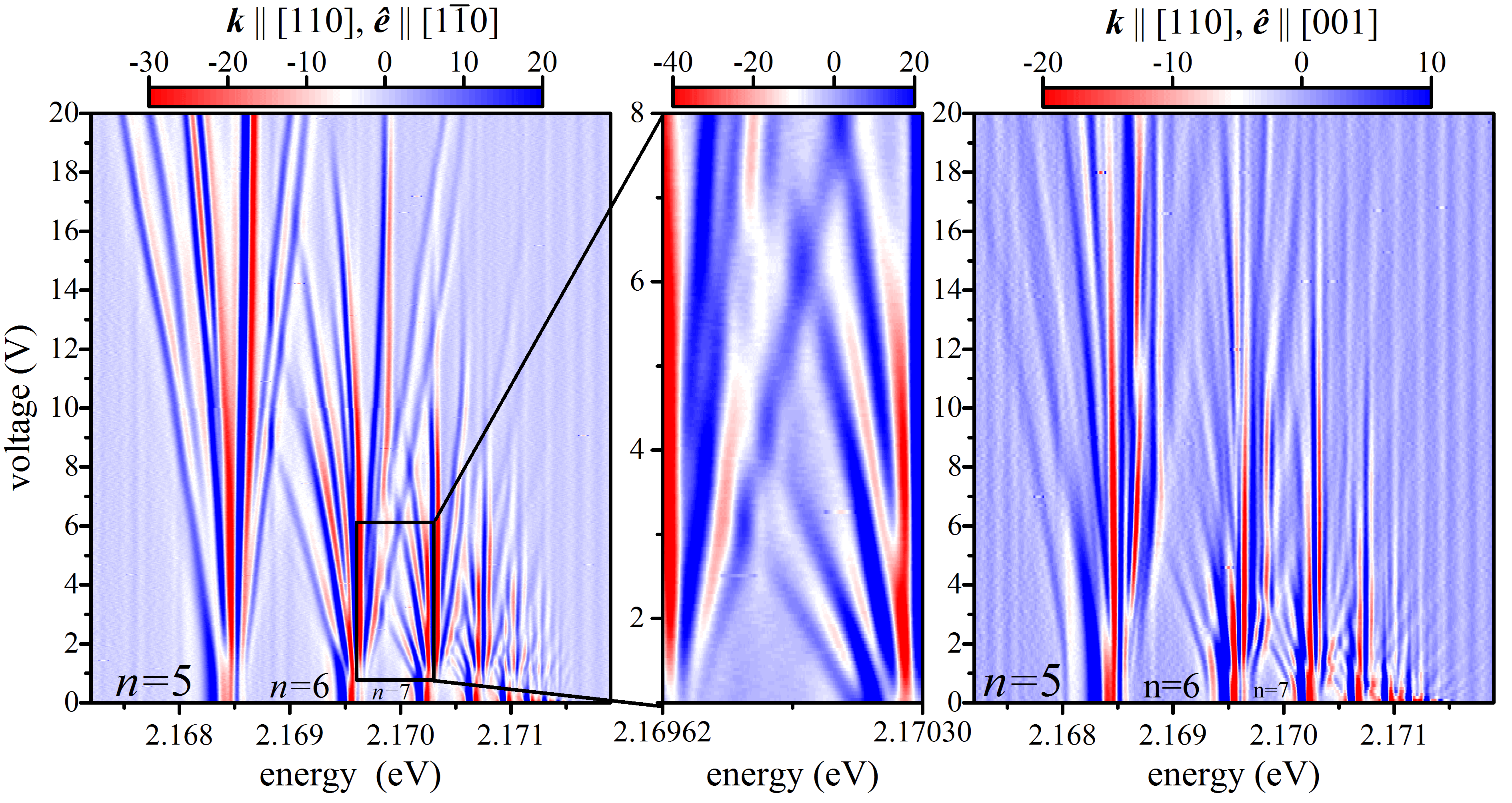}
\caption{Contour plot of the second derivatives of absorption spectra versus applied voltage. The spectra were recorded by white light excitation at $T = 1.3$~K on a Cu$_2$O crystal slab with $[110]$ orientation. The light was polarized along the $[1\bar10]$ direction in the left panel and along the $[001]$ direction in the right panel. The mid panel shows a close-up of the section of the contour plot in the left panel, marked there by the solid box, in order to highlight the anticrossing at the first resonance involving the state of the $n$ = 6 multiplet showing the strongest field dispersion to higher energies and the state of the $n$ = 7 multiplet with the strongest dispersion to lower energies. The weak equidistant vertical stripes in the left and right panel are artifacts of taking the 2nd derivative.}\label{Fig3-efield-spectra}
\end{figure*}

The exciton energies in zero external field can be described well by the quantum defect formalism of Eq.~\eqref{qdef}~\cite{PhysRevB.93.075203}. We recall that in Cu$_2$O the origin of the quantum defect is quite different from that in many-electron Rydberg atoms and is related with the complex valence band structure: The separation between the topmost $\Gamma_7^+$ and the closest $\Gamma_8^+$ valence subbands is on the same order of magnitude as the exciton Rydberg energy $\mathcal R$. Hence, the mixing of these bands via the off-diagonal elements of the Luttinger Hamiltonian gives rise to the deviation of the excitonic series from a hydrogenic one and to the fine structure of the exciton energy spectrum. Moreover, the exchange interaction between the electron and the hole also contributes significantly to the exciton state splitting and mixing~\cite{PhysRevB.23.2731,PhysRevLett.115.027402,PhysRevB.93.195203,efield}. For excitons with principal quantum numbers $n\gtrsim 4$ estimates show that the mixing of the valence subbands can be treated perturbatively and the effective Hamiltonian can be recast in the following form~[cf.~\cite{PhysRevB.93.075203,efield}]:
\begin{multline}
\label{H:eff}
\mathcal H = \frac{p^2}{2\mu} - \frac{e^2}{\varepsilon r} + \mathcal A p^4 +\\
 \frac{e^2}{\varepsilon r^3} \left[\mathcal B_e(\bm l\cdot \bm s_e) + \mathcal B_h(\bm l\cdot \bm s_h)\right] + \\
\mathcal C\delta(\bm r) (\bm s_e \cdot \bm s_h) + \mathcal H_{cubic} + \mathcal H_{sd}.
\end{multline}
Here and through the rest of the manuscript, we consider excitons with center of mass wavevector $\bm K=0$. In Eq.~\eqref{H:eff} the first two terms, subsequently denoted in short by $\mathcal H_0$, describe the standard hydrogen-like problem with the relative motion momentum $\bm p$, the relative motion coordinate $\bm r$, the electron charge $e$ and the dielectric constant $\varepsilon$. The quartic term $\mathcal Ap^4$ describes the nonparabolicity of the kinetic energy with $\mathcal A$ being a constant. The parameter $\mathcal A$ can be estimated in the spherical approximation of the Hamiltonian via the Luttinger parameter $\gamma_2$ as $\mathcal A \approx 2\gamma_2^2/(m_0 ^2\Delta)$, where $m_0$ is the free electron mass and $\Delta$ is the splitting between the $\Gamma_8^+$ and $\Gamma_7^+$ bands. The second line of Eq.~\eqref{H:eff} describes the spin-orbit interaction, where $\bm l=\hbar^{-1}[\bm r\times \bm p]$ is the angular momentum operator, $\bm s_e$ ($\bm s_h$) are the electron (hole) spin-$1/2$ operators acting on the basis functions of the $\Gamma_6^+$ and $\Gamma_7^+$ representations, respectively, and $\mathcal B_e$ ($\mathcal B_h$) are constants. In the third line, the term $\mathcal C\delta(\bm r) (\bm s_e \cdot \bm s_h)$ describes the short-range electron-hole exchange interaction with the exchange constant $C$, and the additional terms $\mathcal H_{cubic}$ and $\mathcal H_{sd}$ account for the cubic anisotropy and the $S-D$-exciton mixing, respectively. In the following the contributions $\mathcal H_{cubic}$ and $\mathcal H_{sd}$ are neglected. The Hamiltonian~\eqref{H:eff} produces quantum defects that are in reasonable agreement with the experiment. To that end we first solve the hydrogenic problem and obtain $\Psi_{nlm}(\bm r)$, which are the eigenfunctions of $\mathcal H_0$. The remaining terms are treated perturbatively. To that end, we introduce the Hamiltonian that is within the applied approximations the extension of the Hamiltonian beyond the hydrogen model:
\begin{equation}
\label{Hd}
\mathcal H_d =\mathcal A p^4 +
 \frac{e^2}{\varepsilon r^3} \left[\mathcal B_e(\bm l\cdot \bm s_e) + \mathcal B_h(\bm l\cdot \bm s_h)\right] + 
\mathcal C\delta(\bm r) (\bm s_e \cdot \bm s_h),
\end{equation}
and is responsible for the quantum defect appearance.
Retaining first-order perturbation theory in $\mathcal H_d$ the contributions leading in the inverse principal quantum number we arrive at~\cite{ll4_eng}
\begin{subequations}
\label{defects}
\begin{align}
&\langle p^4\rangle_{nlm} = \left(\frac{\hbar}{a_B}\right)^4 \frac{4}{n^3(l+1)}, \label{non:par}\\
&\left\langle \frac{1}{r^3} \right\rangle_{nlm} = \frac{1}{a_B^3} \frac{1}{n^3l(l+1/2)(l+1)},\\
& \langle \delta(\bm r)\rangle_{nlm} =  \frac{\delta^K_{l,0}}{\pi} \frac{1}{a_B n^3},
\end{align}
\end{subequations} 
where $\langle \ldots \rangle_{nlm}$ denotes the quantum-mechanical average of the corresponding quantity over the exciton state $\Psi_{nlm}(\bm r)$, $a_B = h^2\varepsilon/(\mu e^2)\approx1.11$~nm is the exciton Bohr radius for the $P$-excitons, and $\delta^K_{a,b}$ is the Kronecker $\delta$-symbol. 

The analysis of the experimental data shows that the exciton quantum defects for large $n$ quickly converge to constant values for fixed angular momentum $l$. This observation is supported by Eqs.~\eqref{defects}, which show that each contribution to the energy deviation of the levels within a multiplet from the simple hydrogen formula scales as $1/n^3$, in agreement with the quantum defect description, Eq.~\eqref{qdef}. Furthermore, we found experimentally that with increasing $l$ the quantum defects drop continuously from finite positive values towards zero, again, in full agreement with Eqs.~\eqref{defects}. The $S$-excitons demonstrate the largest quantum defect $\delta_{n,l=0} \approx 0.65$, for the $P$-excitons it is about $0.34$ and for the $D$- and $F$-excitons it is reduced to $0.18$ and $0.12$, respectively, in the large-$n$ limit. Theoretical estimates of the non-parabolicity contribution to the quantum defects, Eq.~\eqref{non:par}, for $\gamma_2=0.8$~\cite{PhysRevLett.115.027402,PhysRevB.93.195203} yield $\delta_{n,l=0}=0.87$, $\delta_{n,l=1}=0.43$, $\delta_{n,l=2} =0.28$ and $\delta_{n,l=3} = 0.21$ in reasonably good agreement with experiment. It supports our conjecture that the dominant contribution to quantum defects arises from the $p^4$ nonparabolic term in the valence band dispersion.

\begin{figure}[h]
\includegraphics[width=\columnwidth]{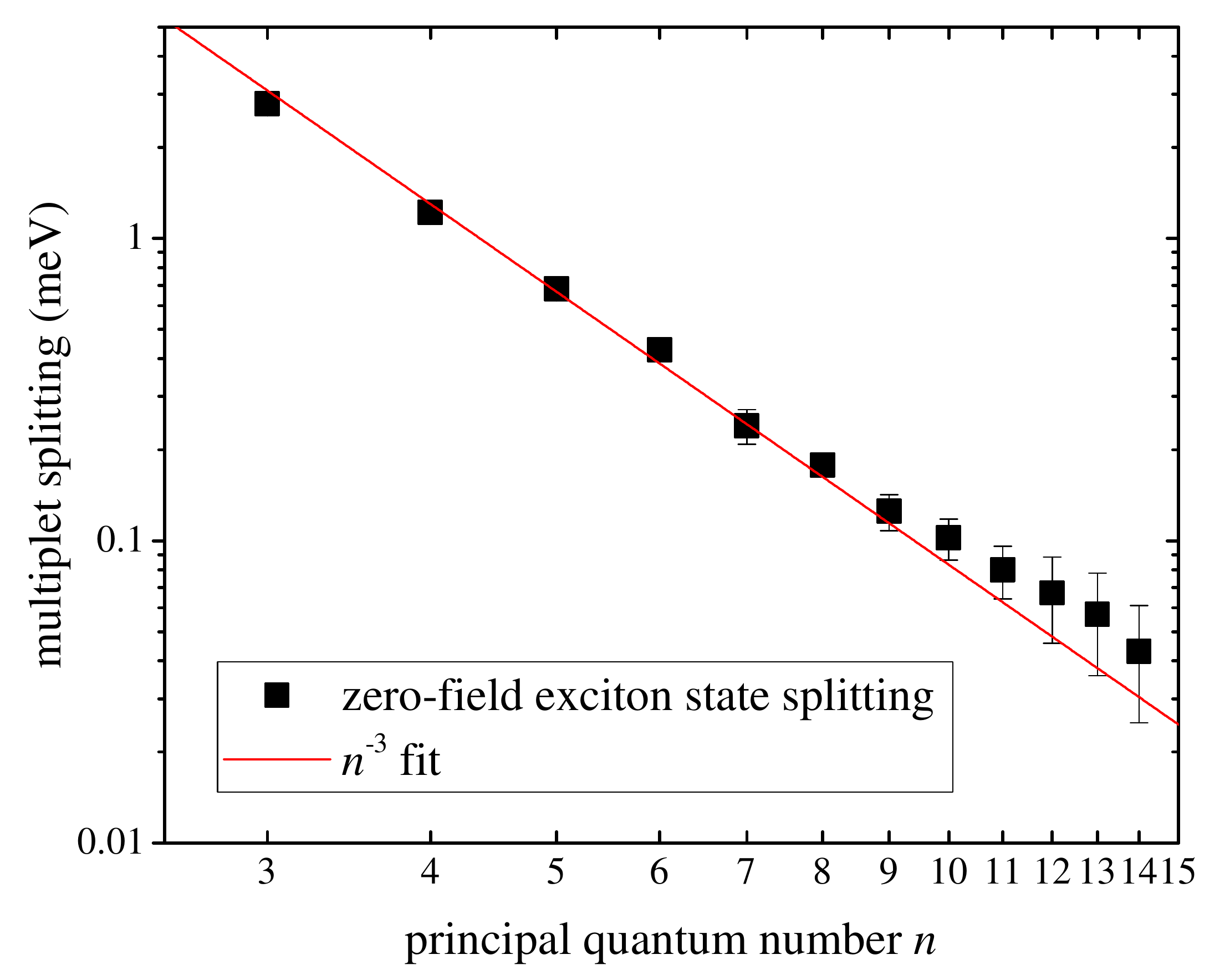}
\caption{Energy spread of the exciton states within a multiplet with fixed principal quantum number versus $n$ in a double logarithmic representation. The line gives a fit to the data after a $n^{-3}$ scaling law, following the theoretical considerations in the text, see Eq.~\eqref{defects}.} \label{Fig4-multipletsplitting}
\end{figure}

The consequence of the quantum defect series for a particular principal quantum number is that the associated states are spread over a finite energy range with the low (high) angular momentum states on the low (high) energy side, as clearly seen in the spectra in Fig.~\ref{Fig2-bfield-spectra-2ndderivative} and in Fig.~\ref{Fig3-efield-spectra}. The widths of these multiplets were determined from the data in zero field and also in weak external fields, applied to optically activate excitons that are dark without field or to enhance their visibility so that their energies could be extrapolated to zero field with high accuracy. The widths of the multiplets are plotted in Fig.~\ref{Fig4-multipletsplitting} as function of $n$ in double-logarithmic representation and show a strong drop with increasing principal quantum number from $3$ meV for $n=3$ to less than 0.1 meV for $n$ exceeding 10. The data follow the $1/n^3$ scaling law expected from the quantum defect description and corroborated by the calculations according to Eqs.~\eqref{defects}, as shown by the red line fit.

The definition of the Rydberg exciton regime is to some extent arbitrary. We take it here as the regime in which the principal quantum number exceeds $n=5$, because the exciton extension given by twice the average exciton radius exceeds then 100~nm, which is roughly two orders of magnitude larger than the ground state wave function extension. Correspondingly the data in Fig.~\ref{Fig4-multipletsplitting} and also in all subsequent scaling plots were fitted for $n \geq 6$, but the plots of the fits were extended also towards lower $n$ in case corresponding data were available like for the multiplet splitting just discussed.

\subsection{Scaling in magnetic field}\label{sec:magn}

The magnetic field behavior of the excitons in the Rydberg regime has been assessed in detail in Refs. \cite{bfield,Aszmann:2016aa}. Entering this regime or $n \gtrsim 6$, the density of states becomes so large that a precise assignment of the observed states is difficult, in particular, because here avoided crossings and therefore state mixings dominate. Hence statistical methods have been applied by analyzing the distribution of energy separations between nearest exciton levels at fixed transition energies. For $n > 6$ clear signatures of quantum chaos were found due to the dominance of anticrossings as can be seen also from the spectra in Fig.~\ref{Fig2-bfield-spectra-2ndderivative} in the energy range above about $2.170$~eV. Here we focus on different features: (i) an estimate of the size of the excitons, and (ii) the magnetic field induced crossing of states with adjacent $n$.

\textit{Exciton size evaluation.} So far, the Rydberg exciton size has been determined somewhat ``indirectly'' from their principal quantum number $n$ using the formula for the average radius of an orbital in the hydrogen model. This average radius is given by~\cite{ll3_eng}
\begin{equation}
\label{extension}
\langle r\rangle_{nlm}  = \frac{a_B}{2} \left[ 3n^2 - l \left( l+1 \right) \right] \approx \frac{3 a_B}{2} n^2  
\end{equation}
with the approximation being well suited for $n \gtrsim 6$ for the $P$-excitons. The magnetic field introduces an independent length scale given by the magnetic length, $\ell_c = \sqrt{\hbar c/ eB}$ characterizing the extension of the magnetic confinement potential, which competes with the Coulomb interaction: At low magnetic fields, the main $P$-exciton lines in Fig.~\ref{Fig1-bfield-spectra} show a rather weak dependence on magnetic field, corresponding to a diamagnetic shift $\propto B^2$, which changes with increasing field to a stronger shift that can be roughly approximated by a $B$-linear dependence. In order to assess this transition theoretically, it is sufficient to present the effective Hamiltonian of the exciton in the magnetic field $\bm B\parallel z$ in the form: 
\begin{equation}
\label{HB:simple}
\mathcal H_B = \mathcal H_0 + \frac{\hbar e B}{2\mu c}l_z + \frac{\hbar^2e^2B^2}{8\mu c^2} (x^2+y^2).
\end{equation}
The terms responsible for the quantum defect are not of importance here. We also disregard the electron and hole Zeeman spin splittings. For strong magnetic fields one can neglect in first approximation the Coulomb interaction and approximate the states by the electron-hole Landau levels. More precisely, even at strong magnetic fields the Coulomb interaction provides bound magnetoexciton states with the binding energy scaling as ${\mathcal R}\ln(a_B/\ell_c)$~\cite{ll3_eng}. With increasing field the changeover from an exciton behavior dominated by the Coulomb interaction at low fields to a behavior with dominant magnetic confinement at high fields, where Landau levels are formed, occurs. 

Correspondingly, an exciton resonance in the optical spectrum transforms to a good approximation into a transition between electron and hole Landau levels. Roughly, $P$-excitons with principal quantum number $n$ transform into a transition between Landau levels with quantum number $n$~\cite{Yafet1956137,PhysRev.188.1294}. The extension of these Landau levels in real space is given by
\begin{equation}
\label{lcn}
\ell_{c,n}  \approx \ell_{c} n^{1/2} = \sqrt{ \frac{\hbar c n}{eB} } 
\end{equation}
for not too small $n$. The transition takes place for 
\[
\ell_{c,n} \sim \langle r \rangle_{nlm},
\]
i.e., where the Landau level extension is about equal to the Coulomb extension, which can be achieved by increasing the magnetic field to a particular crossover field strength $B_{c,n}$. From this crossover field one obtains $\ell_{c,n} = 25.6~$nm$ \sqrt{n} / \sqrt{B_{c,n} [\mathrm T]}$, where $B_{c,n}$ should be inserted in units of Tesla. In the experiment we can determine $B_{c,n}$ as the field strength at which the quadratic field dependence of the $P$-exciton energies changes into a linear one. Note that the crossover fields and the resulting Landau level extensions can be only approximately determined in that way, but they are nevertheless sufficient to test their scaling with $n$. From the considerations above we expect a scaling with $n^{-3}$ for $B_{c,n}$.

\begin{figure}[h]
\includegraphics[width=\linewidth]{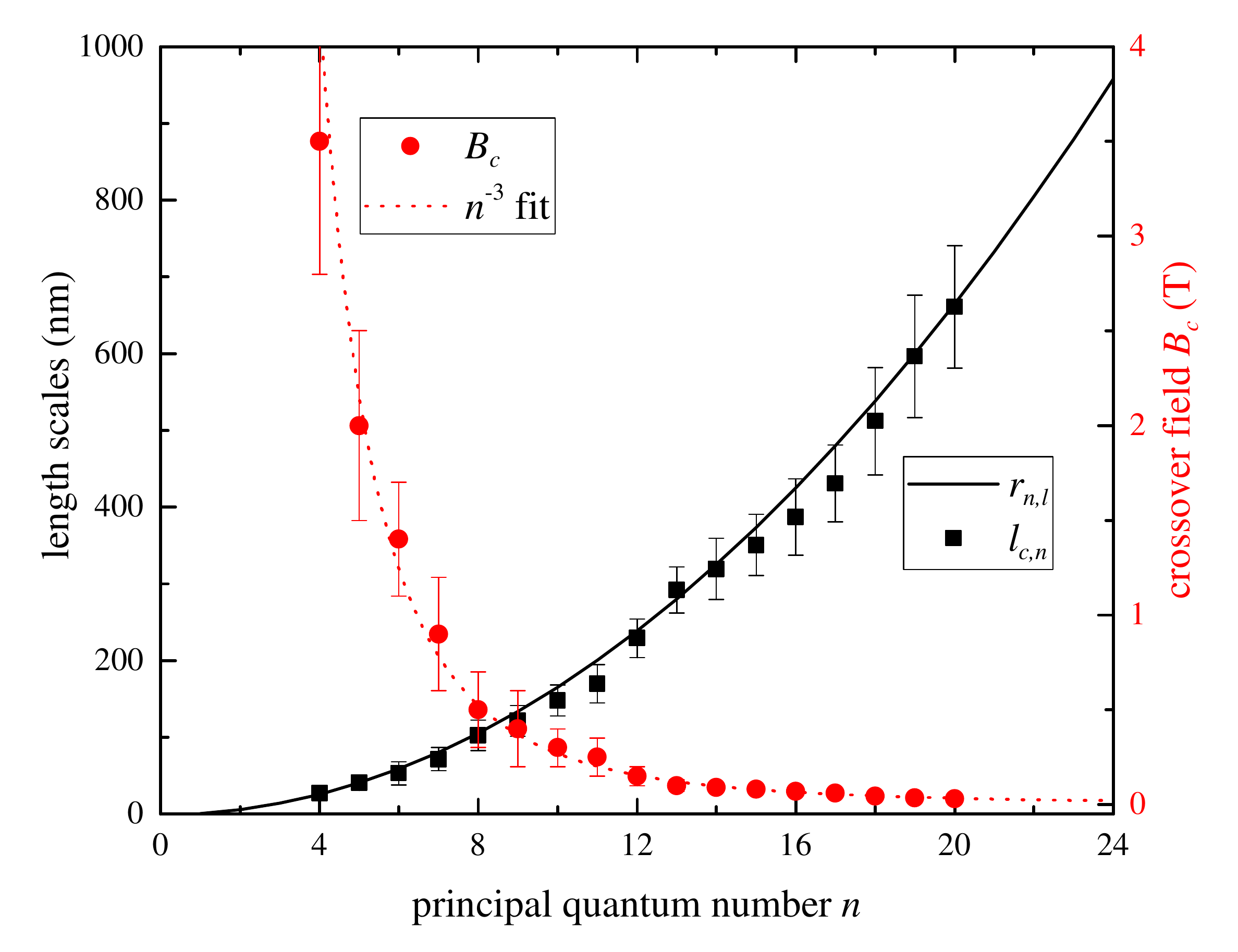}
\caption{Red circles: crossover field strength $B_{c,n}$ from diamagnetic to Landau level-like behavior as function of principal quantum number. The red dotted line is a fit to the data using a $n^{-3}$ dependence. Black squares: Landau level radius $\ell_{c,n}$, Eq.~\eqref{lcn}, estimated from $B_{c,n}$ as function of $n$. The black solid line gives the average radius of the orbital according to the hydrogenic formula, Eq.~\eqref{extension}.} \label{Fig5-lengthscales}
\end{figure}

From the spectra it is confirmed that the transition to Landau-level like behavior occurs at continuously decreasing magnetic fields with increasing principal quantum number. In the high field regime, where also quantum chaos was detected for $n > 6$ no single strong transition is observed for a particular $n$, but bunches of lines which cluster around the Landau level transitions appear. Technically, to determine the crossover field we have taken the center of the line multiplet around a Landau transition and have extrapolated its energy linearly to lower fields, so that we could determine the crossing point $B_{c,n}$ with the quadratic diamagnetic shift of the $P$-exciton line. The $B_{c,n}$ determined in that way are shown by the red circles in Fig.~\ref{Fig5-lengthscales}, demonstrating the strong drop with increasing $n$. The dotted line gives a $n^{-3}$ fit to the data from which excellent agreement is seen.

From these fields $\ell_{c,n} ( B_{c,n} )$ can be calculated. The Landau level extensions determined in that way are shown in Fig.~\ref{Fig5-lengthscales} by the black squares and compared to the average exciton radius $r_{n,l}$ according to the hydrogenic formula. For example, for $n=6$ the hydrogenic formula~\eqref{extension} gives an average radius of 60~nm. The changeover field strength, on the other hand, is $1.3 \pm 0.3$~T, from which we obtain $\ell_{c,6}(B_{c,6}) \approx 55\pm 8$~nm which are surprisingly close values. For all other $n$ we find similarly good agreement between $\ell_{c,n}$ and $\left( r_{n,l} \right)$, confirming the expected $n^2$ scaling of the wave function extension: $\ell_{c,n} \propto (n/B_{c,n})^{1/2} \propto \sqrt{n/n^{-3}}$. Note that the inaccuracy in determining $B_{c,n}$ translates through the square-root connection in moderate inaccuracies of the Landau level extension $\ell_{c,n}$ that are indicated by the error bars. For high $n$ sufficient accuracy was achieved by recording the spectra in 5~mT steps which in combination with the steeper slope of the Landau level transitions facilitates the separation from the low-field behavior. While the observed agreement with the hydrogenic formula may be expected, it also validates the applied exciton description using, for example, a uniform dielectric screening over large length scales. It also reassures the huge extension of the highly excited Rydberg exciton, for which we estimate for $n=20$ an average radius of about 0.7~$\mu$m.

\emph{Magnetic field-induced crossings.} For the field-induced resonances we focus on the first one between exciton states belonging to the multiplets with principal quantum numbers $n$ and $n+1$, as shown in detail in the bottom panel of Fig.~\ref{Fig2-bfield-spectra-2ndderivative} for field strengths up to $2$ T.  At these resonances we observe systematic crossings, in contrast to the general trend of anticrossings in the spectra. From lower to higher energies we clearly see that the field strength, $B_r$, at which these crossings occur shifts strongly to lower values. This can be expected from (i) the reduced splitting between the exciton states at zero field, (ii) the enhanced field-induced splitting of each of them involving larger angular momenta, and (iii) the stronger diamagnetic shift of each multiplet for higher $n$.

\begin{figure}[h]
\includegraphics[width=\linewidth]{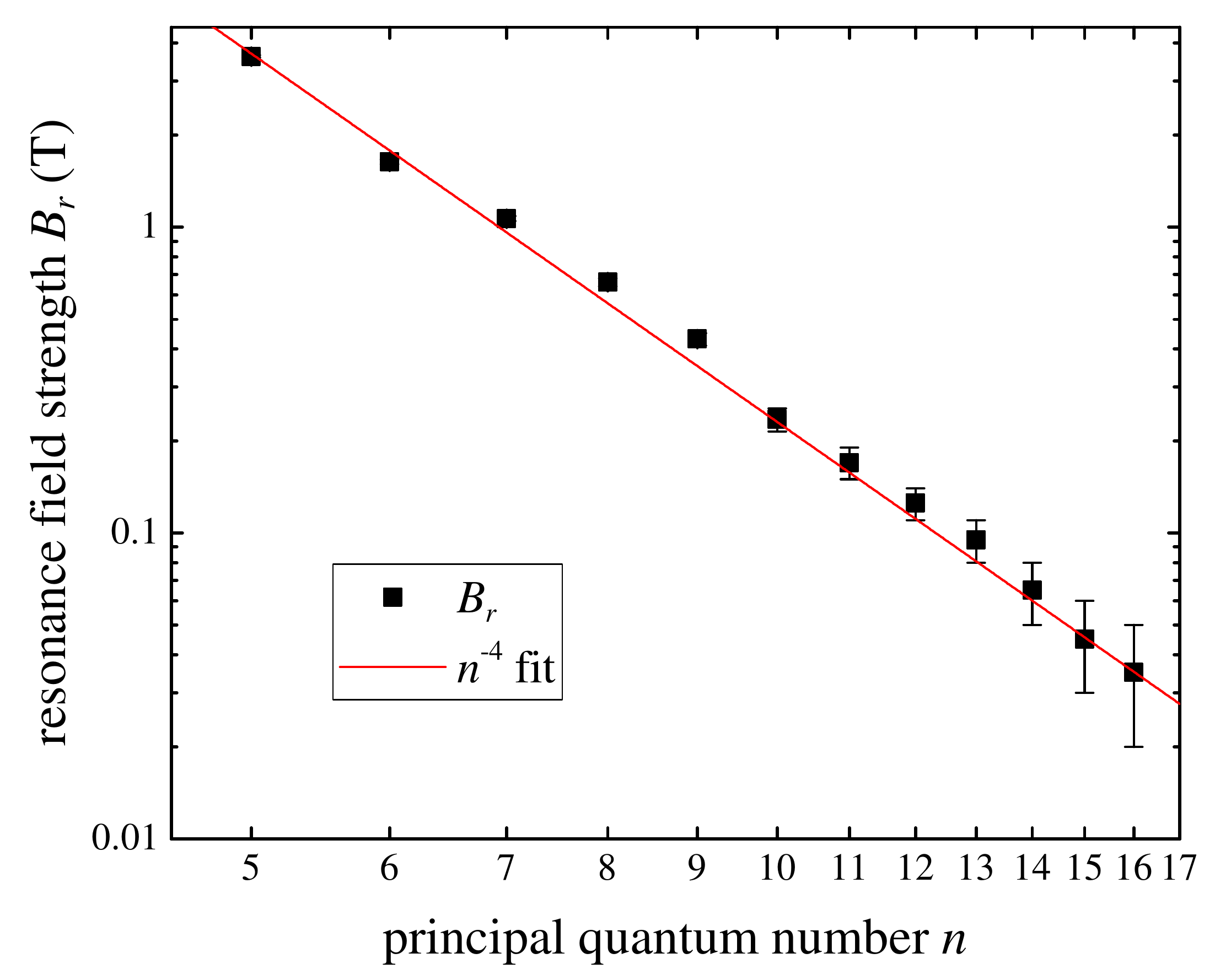}
\caption{Resonant magnetic field strength, $B_r$, at which the first crossing between states belonging to the exciton multiplets $n$ and $n+1$ occurs, as function of the principal quantum number $n$ in a double logarithmic presentation. Symbols give the experimental data, the line is a fit to these data following a $n^{-4}$ law, Eq.~\eqref{Bc}.}\label{Fig6-resonantmagneticfields}
\end{figure}

From the data we can determine the resonant field strengths $B_r$ versus the principal quantum number. This dependence is shown in Fig.~\ref{Fig6-resonantmagneticfields} using a double-logarithmic representation. The resonant fields decrease from $B_r=2$~T for $n=6$ to $B_r=0.04$~T for $n=16$.
Fitting the observed data for the Rydberg exciton regime with a power law form reveals a dependence as $n^{-4}$, unlike for Rydberg atoms, where one finds a $n^{-6}$-dependence~\cite{gallagher2005rydberg}.

This scaling can be well understood by applying the simplified hydrogen-like model, which is to a good approximation justified because of the quite small impact of the quantum defect on the large-$n$ Rydberg states, as already demonstrated by the $n^{-3}$ scaling of the multiplet width in Fig.~\ref{Fig4-multipletsplitting}. Using atom-like excitonic units, i.e., $2\mathcal R = \mu e^4/(\varepsilon^2\hbar^2)$ as the unit of energy, the Bohr radius $a_B$ as the unit of length, and $\hbar c/(ea_B^2)$ as the unit of magnetic field, and 
\begin{equation}
\label{r2}
\langle r^2\rangle_{nlm} = \frac{n^2}{2}[5n^2+1-3l(l+1)] \propto n^4
\end{equation}
for sufficiently large $n$, we obtain for the level energies to a good approximation:
\begin{equation}
\label{Enl:B}
E_{nl} \approx -\frac{1}{2n^2} + \frac{m B}{2} + A_l n^4 B^2.
\end{equation}
Here $A_l$ is a constant that depends on the angular momentum $l$ of the studied level. Neglecting the diamagnetic shift $\propto B^2$, we obtain for the resonance field of the sublevels $n,l=n-1,m=n-1$ and $n+1, l=n, m=-n$, which are involved in the crossing (corresponding to the states with the maximal $m$ of the $n$th multiplet and with the minimal $m$  of the $n+1$st multiplet), in perturbation approach:
\begin{equation}
\label{Bc}
B_r(n) \propto \frac{1}{n^4}.
\end{equation}
It is noteworthy that for the crossing of these states the diamagnetic contribution is vanishingly small, because $A_l n^4 B_r^2(n)\propto 1/n^4 \ll mB \propto 1/n^3$ in Eq.~\eqref{Enl:B}.  This is different from the atomic physics case where one, as a rule, considers states with magnetic quantum numbers, $m=0$ or $1$, as only these states can be observed in single photon absorption out of an $s$-state. Correspondingly, in the atomic case the paramagnetic term $\propto mB/2$ in Eq.~\eqref{Enl:B} is not important and the diamagnetic term $\propto A_l n^4 B^2$ dominates~\cite{gallagher2005rydberg}. For atoms one finds therefore a $n^{-6}$ dependence of the resonance field strength. 

\subsection{Electric field}

Let us turn now to the scaling of exciton properties in electric field. As outlined in Sec.~\ref{sec:exper}, we recorded spectra for two different configurations, taken on a $[110]$ oriented crystal. For the exciting light propagating along the same $[110]$-direction the linear light polarization was chosen either along $[1\bar10]$ (left panel of Fig.~\ref{Fig3-efield-spectra}) or along $[001]$ (right panel of the same figure). In the first case the quadrupolar transitions are forbidden so that it is easy to follow the dispersion of the $P$-excitons, which is the first major point of this subsection. In the latter case, they are allowed, so that $S$- and $D$-excitons appear, allowing a more comprehensive insight into the different states within an exciton multiplet~\cite{efield}. This is exploited here to determine the fields at which states from different multiplets come in resonance, representing the second major point of this part. The same configuration was also used to study the polarizability of the $S$-excitons which are well separated on the low energy flanks of the $P$-excitons. Finally we also address exciton ionization processes, for which we use again the first configuration (without quadrupolar transitions) to determine the ionization field strength as well as the linewidth of $P$-excitons.

\emph{Scaling of $S$- and $P$-exciton polarizability.} For the $P$-exciton dispersion, we have to distinguish between the low-$n$ and high-$n$ states. For low $n$ the effect of the quantum defect, namely the lifting of level degeneracy, is particularly relevant. In electric field, this leads to the observation of a quadratic Stark effect for the non-degenerate states, because in centrosymmetric crystals any non-degenerate excitonic state has no electric dipole moment without electric field application. The dipole moment is induced by the electric field which subsequently orients it, leading to the quadratic energy shift. This is in contrast to the linear Stark effect obtained by degenerate perturbation theory for a multiplet of levels each having the same energy. In this case, the degenerate states in the multiplet that become coupled by the electric field are linearly combined such that an electric dipole moment is established. This dipole moment only has to be oriented by the field leading to linear energy shifts with increasing field. This situation is to a good approximation relevant for the high $n$-range. Strictly speaking, the exciton differentiation between small and large values of the principal quantum number as well as between non-degenerate and degenerate states depends on the experimental resolution: For each $n$, independent of its value, there is a range of small fields in which a quadratic Stark shift occurs due to level splitting in the crystal. Roughly, this is the range in which the associated quantum defect exceeds the field induced energy shift. This range of quadratic Stark effect decreases, however, strongly with increasing $n$.

\begin{figure}[h]
\includegraphics[width=\linewidth]{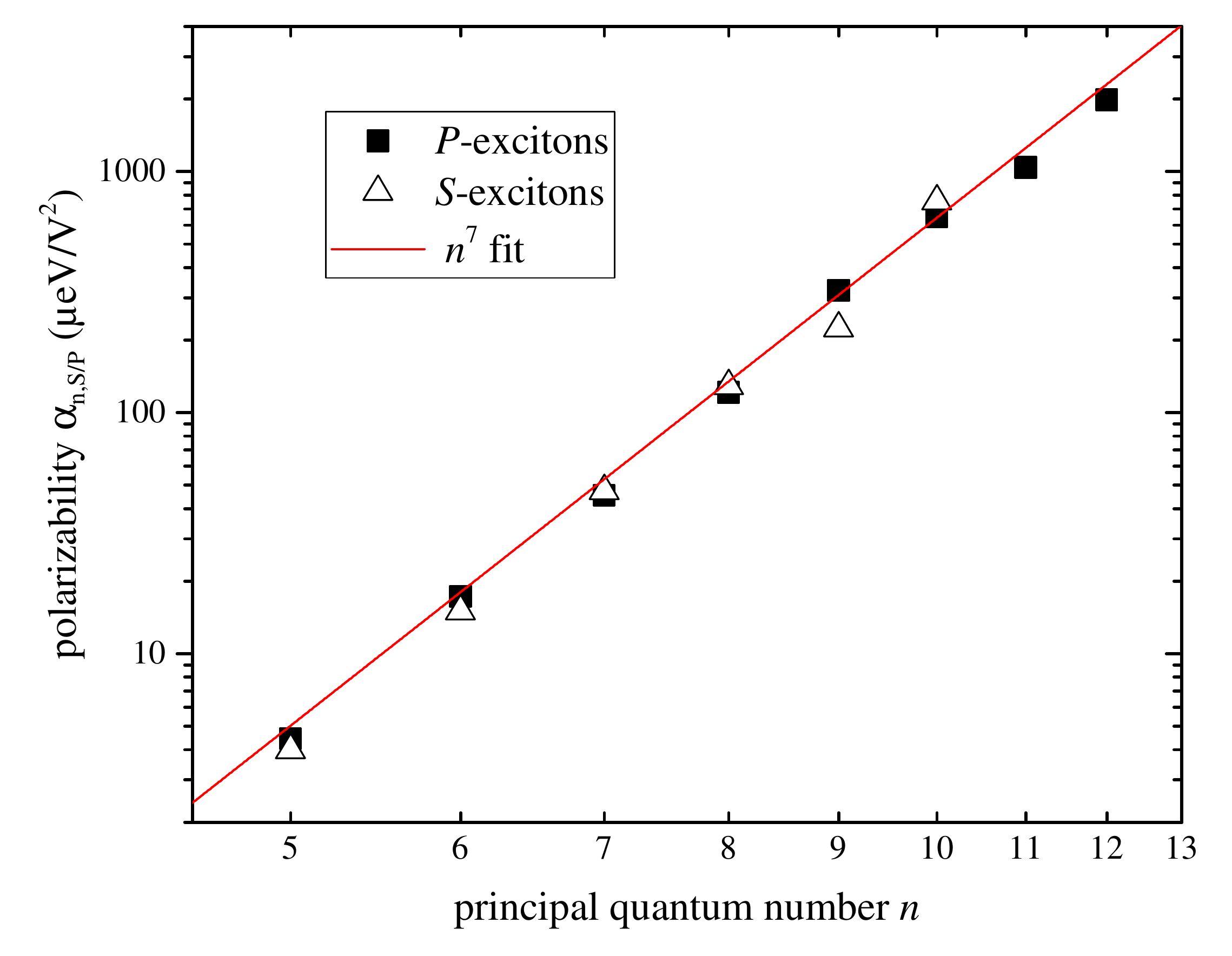}
\caption{Polarizability $\alpha_{n,l}$ of the $S$- (open triangles) and $P$-excitons (closed squares) versus principal quantum number $n$. The line is a fit to the data following the expected $n^7$ scaling law, Eq.~\eqref{pol:scaling}.}\label{Fig7-polarizabilitySP}
\end{figure}

The quadratic Stark effect is described by
\begin{equation}
\label{quadratic}
\Delta E_{nl}^{(2)} = - \alpha_{nl} F^2,
\end{equation}
where $\alpha_{nl}$ is the polarizability of the exciton state $(n,l)$. Although a quadratic Stark effect is already expected for the hydrogenic model with $\alpha_{nl} \propto n^6$ due to the field induced mixing of the multiplets with different principal quantum numbers~\cite{ll3_eng}, the presence of the quantum defect leads to an important change of the polarizability scaling with $n$: Mixing becomes possible for states within a multiplet of particular $n$, which are split due to the $\mathcal H_d$ part of the Hamiltonian, Eq.~\eqref{Hd}. The scaling relation of the polarizability can be understood making use of second order perturbation theory to evaluate the quadratic Stark shift: Taking into account the dipole moment operator matrix elements between neighbouring states with the same $n$ that scale as $e\langle r\rangle_{nlm} \propto n^2$, Eq.~\eqref{extension}, and the scaling of the energy gap between these states $\propto 1/n^3$, see Fig.~\ref{Fig4-multipletsplitting} and Sec.~\ref{sec:zero}, the polarizability scales as
\begin{equation}
\label{pol:scaling}
\alpha_{nl} \propto n^7,
\end{equation}
in accordance with the behavior typically observed for Rydberg atoms~\cite{PhysRevA.62.042703,RevModPhys.82.2313}.

Here, we observe for the $P$-excitons in Fig.~\ref{Fig3-efield-spectra} a quadratic Stark effect for $n \leq$ 13, in agreement with Eq.~\eqref{quadratic}, while the splitting pattern for the higher-$n$ excitons approaches the linear Stark fan of hydrogen within the experimental resolution. In Fig.~\ref{Fig3-efield-spectra}(a) the shift of the $P$-excitons to lower energies with increasing field can be well resolved and increases drastically with increasing principal quantum number, see the blue-colored feature of lowest energy in each $n$-manifold. From the data we can assess the polarizability $\alpha_{n,P}$ ($l=1$) which is shown by the solid squares in Fig.~\ref{Fig7-polarizabilitySP} as function of $n$ in a double-logarithmic plot. The polarizability increases from about 5 $\mu$eV/V$^2$ for $n$ = 5 to about 2000 $\mu$eV/V$^2$ for $n$ = 12. The solid lines give the fit to these data by a power law scaling with the seventh power of $n$, in accordance with the theoretical expectations, Eq.~\eqref{pol:scaling}, from which a good description of the data is obtained. In Fig.~\ref{Fig3-efield-spectra}(b), due to the chosen polarization configuration, also the $S$-excitons can be observed, which show also a quadratic Stark effect. Deriving also for them the polarizability, which is possible for $n \leq 10$, one obtains the data shown by the open triangles in Fig.~\ref{Fig7-polarizabilitySP}. Within the experimental error no difference in polarizability between $S$- and $P$-excitons can be resolved.

\begin{figure}[h]
\includegraphics[width=\linewidth]{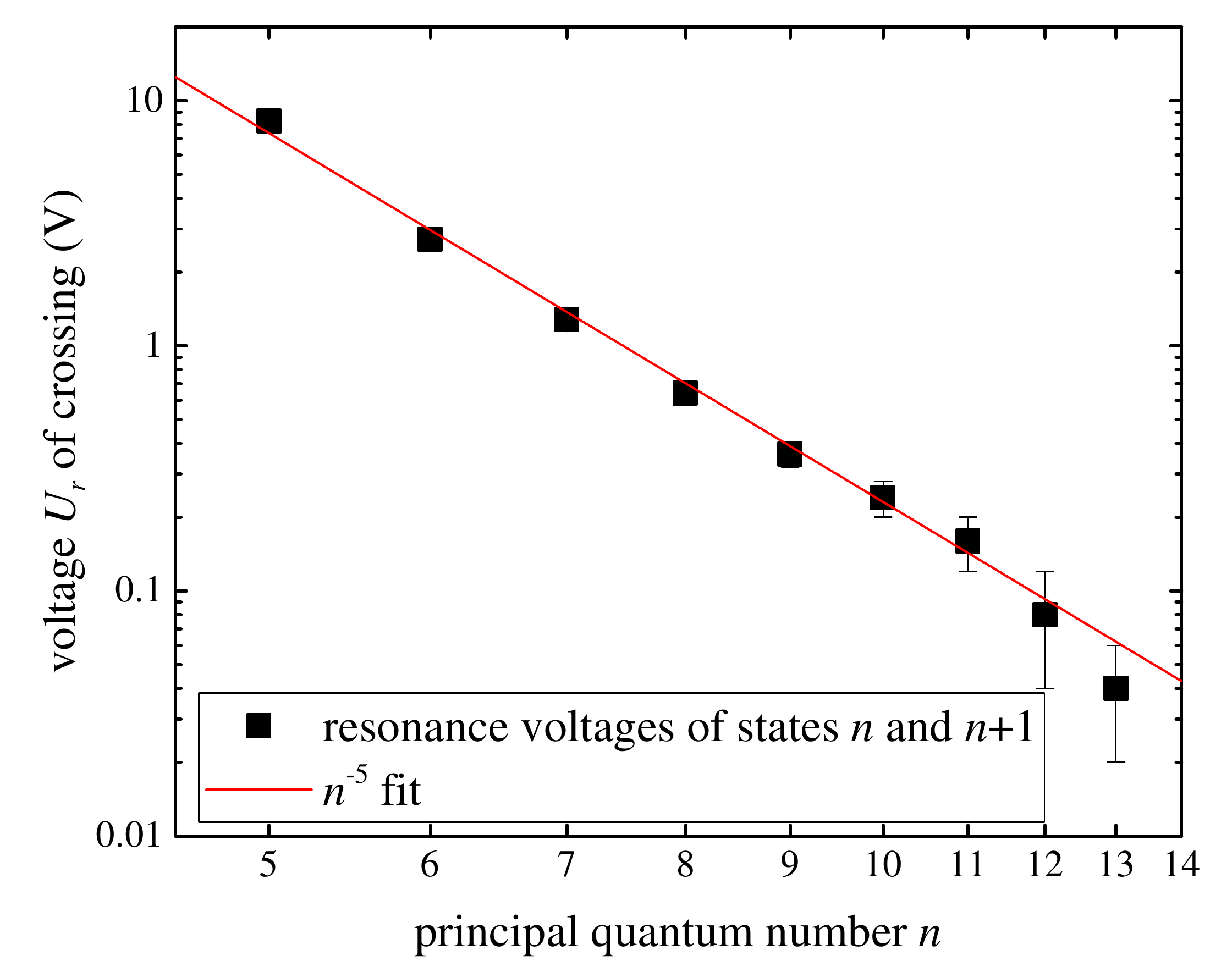}
\caption{Principal quantum number $n$-dependence of the resonance voltage $U_r$ at which the first resonance of levels belonging to the multiplets with principal quantum numbers $n$ and $n+1$ occurs. At these resonances avoided crossings are observed. The red line shows a fit according to a $n^{-3}$ scaling law, Eq.~\eqref{Fc}.}\label{Fig8-resonancevoltages}
\end{figure}

\emph{Scaling of electric field induced resonances.} Next we turn to the first resonances occurring  with increasing electric field between states of adjacent principal quantum numbers. Similar to the magnetic field case, we find also here a strong shift of the resonance voltage $U_r$ to lower values with increasing $n$. Looking at the resonances in closer detail, we find, however, that at these resonances levels do not cross but systematically avoid each other, as shown in the mid panel of Fig.~\ref{Fig3-efield-spectra}. Extrapolating the dispersions, one can determine the resonance voltages which are shown in Fig.~\ref{Fig8-resonancevoltages} in a double logarithmic representation versus $n$. $U_r$ decreases from 8~V for $n=5$ to about 40~mV for $n=13$. The data are in reasonably good accord with a power law scaling like $n^{-5}$ which is in line with the results on Rydberg atoms. Resonance voltage and field strength $F_r$ are connected by a simple proportionality relation for the chosen capacitor geometry by which the field is applied. However, care needs to be exercised in doing this conversion using just the nominal geometric and dielectric parameters, because surface charges, charged defects etc. can lead to depolarization effects in the crystal, so that there may be some discrepancy between the nominally calculated and the actually present field strength by a factor of $3 \ldots 5$, as discussed in detail in Ref.~\cite{efield}.

This behavior can be understood again using the hydrogen-like model, because the quantum defect is not essential for the evaluation of the resonant fields. In the presence of an electric field the simplified Hamiltonian of the exciton reads:
\begin{equation}
\mathcal H_E = H_0 + eFz,
\end{equation}
where we assumed that the field $\bm F$ is applied along the $z$-direction. In first-order perturbation theory:
\begin{equation}
\label{Enl:F}
\frac{E_{nl}}{2\mathcal R} = -\frac{1}{2n^2} + \frac{3}{2}\tilde Fn(n_1-n_2),
\end{equation}
where $\tilde F = Fa_B/(2\mathcal R)$ is the reduced electric field. Here $n_1$ and $n_2$ are the parabolic quantum numbers being non-negative integers with $n=n_1+n_2+|m|+1$. The resonance field strength, $F_r$, can be evaluated from Eq.~\eqref{Enl:F} putting $n_1-n_2 \approx \pm n$ for $n\gg 1$ and for the states with highest and lowest energy of the manifold with the result~\cite{gallagher2005rydberg}:
\begin{equation}
\label{Fc}
F_{r} \propto \frac{1}{3n^5}.
\end{equation}
Note that the power-law dependence is different from the magnetic field case. This is because the electric dipole moment increases as $n^2$~\cite{ll3_eng}, whereas the magnetic field induced splitting is determined by the paramagnetic contribution, which scales linearly with $n$ for the lowest and highest energy states in a multiplet.

\begin{figure}[h]
\includegraphics[width=\linewidth]{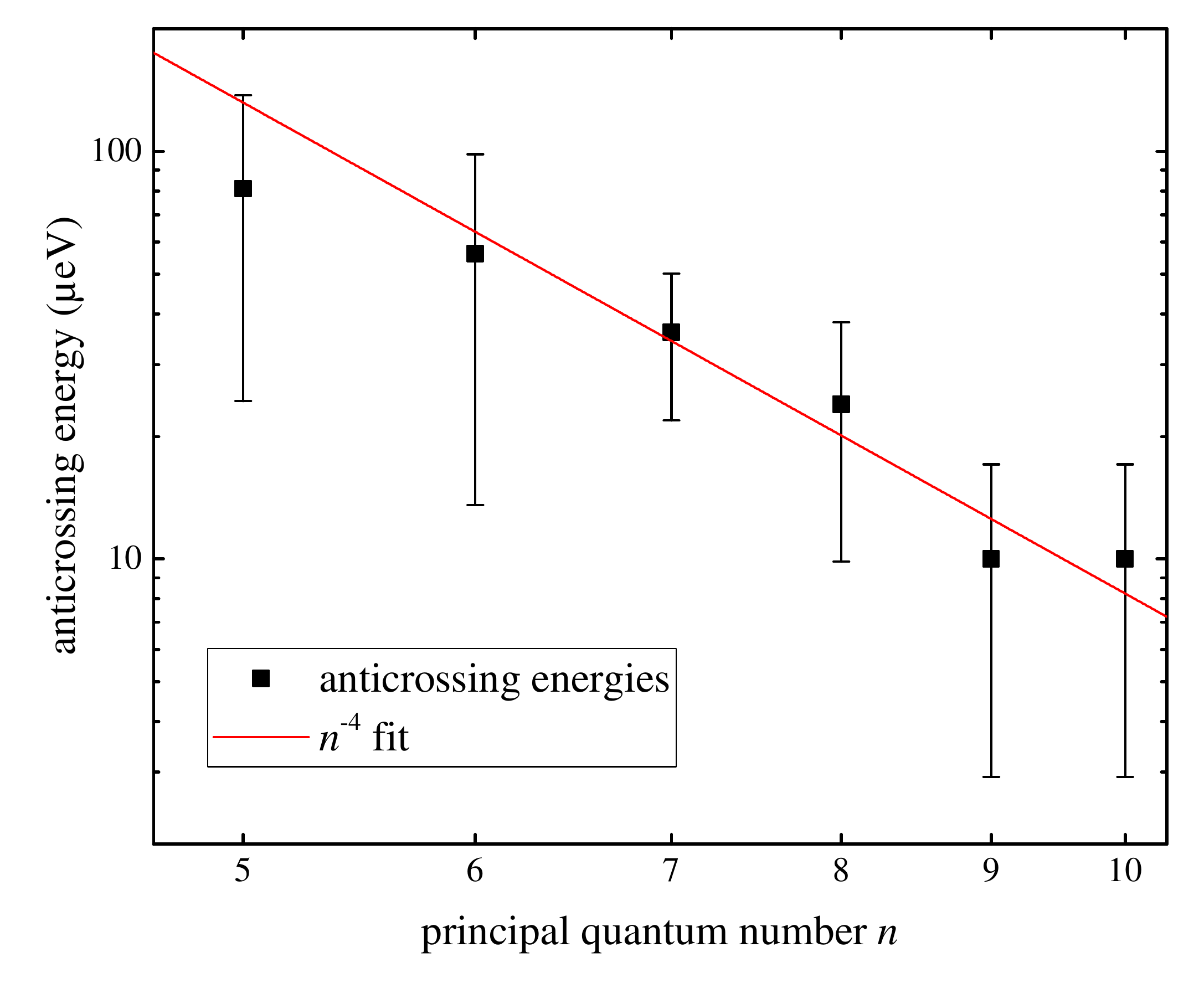}
\caption{Black squares: Dependence of the energy splittings due to anticrossings at the first resonances of states from the exciton multiplets $n$ and $n+1$, induced by the electric field, on the principal quantum number. The red line gives a fit to the data according to a $n^{-4}$ dependence, Eq.~\eqref{anti:stark:sc}.}
\label{Fig9-anticrossingenergies}
\end{figure}

We have also determined the energy splittings at these anticrossings as function of $n$ which are shown in Fig.~\ref{Fig9-anticrossingenergies}. The data show a strong drop with $n$, even though one has to emphasize the splitting magnitude is only slightly larger than the linewidths of the involved transitions, resulting in considerable error bars, despite of which the drop from about 80 $\mu$eV splitting for $n$ = 5 to about 10 $\mu$eV for $n$ = 10 can be determined. The data can be reasonably well described by a $n^{-4}$ fit following also from perturbation theory. In order to estimate the anticrossing energy we follow Refs.~\cite{gallagher2005rydberg,popov} and evaluate the matrix elements of the ``quantum defect'' Hamiltonian~\eqref{Hd} on the eigenfunctions of the hydrogen atom in the parabolic coordinates $\Psi_{nn_1n_2m}(\bm r)$, where $n_1$, $n_2$, and $m$ are the parabolic quantum numbers. The avoided crossing energy reads
\begin{equation}
\label{anti:stark}
\delta E_{anti} = 2\left|\int d\bm r \Psi_{n'n_1'n_2'm}^*(\bm r) \mathcal H_d \Psi_{nn_1n_2m}(\bm r)  \right|.
\end{equation}
For the hydrogenic states with adjacent principal quantum numbers $n'=n\pm 1$ we have~\cite{gallagher2005rydberg}
\[
\int d\bm r \Psi_{n'lm}^*(\bm r) \mathcal H_d \Psi_{nlm}(\bm r) \sim \frac{\mathcal R\delta_{nl}}{n^3}.
\]
Finally, taking into account that 
\[
\sum_l \int d\bm r \Psi_{n'n_1'n_2'm}^*(\bm r)\Psi_{nlm}(\bm r) \propto \frac{1}{n-m},
\]
we obtain the scaling law
\begin{equation}
\label{anti:stark:sc}
\delta E_{anti} \sim \frac{\mathcal R \delta_{nl}}{n^3(n-m)} \propto 1/n^4.
\end{equation}
Using for a rough estimate $\delta_{nl}=0.5$ (Sec.~\ref{sec:zero}) we obtain for the $n=5$ and $n=6$ multiplets an anticrossing energy of about $70$~$\mu$eV in reasonable agreement with the data in Fig.~\ref{Fig9-anticrossingenergies}.

\emph{Crossing vs. anticrossing}. To understand the origin of crossing or anticrossing at the first resonance of states arising from the multiplets with principal quantum number $n$ and $n+1$ in magnetic and electric field, respectively, we have to consider the problem in more detail: Both electric and magnetic field preserve the axial symmetry of the ``hydrogenic'' problem, therefore the $z$-component of the angular momentum or ``magnetic'' quantum number $m$ is a good quantum number. In magnetic field the first crossing occurs between the states with $m=n-1$ (the maximal $m$ in the manifold with $n$) and $m=-n$ (the minimal $m$ in the manifold with $n+1$). The difference of the angular momentum components of these states is
\begin{equation}
\label{Delta:lz:B}
\Delta l_z = 2n+1.
\end{equation}
This is an odd number $1$, $3$, \ldots, so that the parity of excitonic states which cross in magnetic field is different. In the centrosymmetric point group $O_h$ the states are characterized by a definite parity as well and, therefore, states with odd $\Delta l_z$ do not mix. Hence, one may expect that the first crossing in magnetic field is indeed an \emph{allowed} crossing. In electric field the situation is different. One can readily check by perturbative calculations that the first resonance occurs for states with $m=0$, e.g., $|n=2S\rangle + |n=2P_z\rangle$ and $|3S\rangle - |3P_z\rangle$. These states have the same symmetry and, hence, the crossing of these states is \emph{avoided}.

However, one has to be careful when using this line of argumentation to discuss the crossings/anticrossings of \emph{observed} states in the transmission spectra. In fact, the consideration above neglects the symmetry of the two-particle Bloch function of the exciton~\cite{PhysRevLett.115.027402}. All observed excitonic states (in a given geometry and polarization) interact with light and, hence, have the \emph{same} symmetry, $\Gamma_4^-$. For example, the states observed in the geometry where light with $[1\bar10] \parallel y$ polarization propagates along the $[110] \parallel z$ direction, couple to the $y$-component of the electric field and therefore transform as this $y$-coordinate. Hence, at least the light-matter interaction, i.e., the polariton effect~\cite{excitons:RS,PSSB:PSSB2221730104,PhysRevB.94.045205}, can convert allowed crossings into avoided ones. However, the polariton effect is relatively weak in cuprous oxide crystals where the direct interband transitions are forbidden at the $\Gamma$-point of the Brillouin zone, leading to a small oscillator strength of allowed exciton transitions~\cite{NIKITINE1961292,PSSB:PSSB2221730104}. The quantitative measure for the strength of the polariton effect is the longitudinal-transverse splitting, $\hbar\omega_{nlm,LT}$, of the exciton state with quantum numbers $(n,l,m)$. Provided that the longitudinal-transverse splitting is smaller than the non-radiative damping of the excitonic states, 
\[
\hbar\omega_{nlm,LT} < \hbar \Gamma_{nlm},
\]
the light-matter interaction is in the weak coupling regime so that polariton modes are not observable, and also the anticrossings are hidden in the optical spectra.

\begin{figure}[h]
\includegraphics[width=\linewidth]{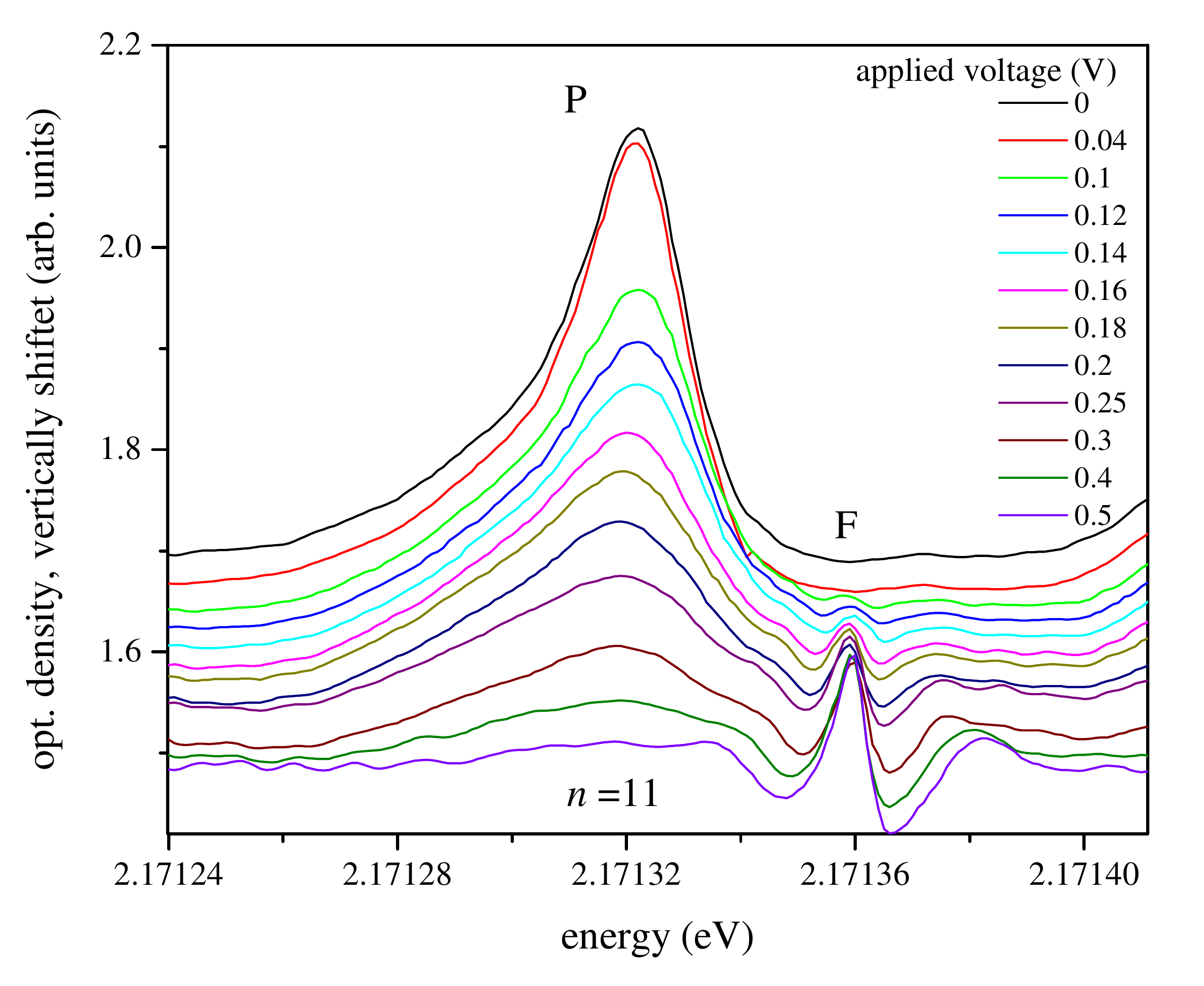}
\caption{High resolution absorption spectra of the $n=11$ exciton multiplet recorded with a frequency stabilized laser for different applied voltages.}\label{Fig10-highresolutionspectrainefieldn=11}
\end{figure}

\emph{Ionization in electric field.} A particularly appealing feature of Rydberg excitons in comparison to Rydberg atoms is the fact that the true high electric field regime, in which the interaction strength with the field exceeds the Coulomb interaction, can be reached easily in the laboratory, while for atoms it is much harder to enter this limit. This is mostly due to the different Rydberg energy in cuprous oxide, $\sim90$~meV, which is more than two orders smaller than the hydrogen atom Rydberg energy. From Fig.~\ref{Fig3-efield-spectra} one sees that for elevated electric field strengths the excitons dissociate: The absorption lines fade away and finally disappear with increasing field. The fields required for dissociation are the smaller, the higher the principal quantum number of the involved exciton state is. A series of spectra recorded for the $n=11$ multiplet is shown in Fig.~\ref{Fig10-highresolutionspectrainefieldn=11}, from which the drop of the $P$-exciton resonance with increasing applied voltage can be quantitatively assessed. Simultaneously with the line drop the width of the resonance increases.

On the high energy side of the $P$-exciton another peak can be seen, which can be associated with the $F$-excitons that can hardly be identified without applied voltage, but become more and more pronounced with increasing voltage. At higher energies even further weak features appear. This increase of visibility is partly related to the transfer of oscillator strength from the $P$ to the $F$ and other higher angular momenta excitons. It is somewhat surprising that the $F$-exciton absorption is still growing even when the $P$-exciton absorption has basically completely vanished. A potential explanation is the action of the centrifugal barrier in the kinetic energy which varies with the angular momentum of the involved state. Due to its action the particles are kept away from the field axis $z$, making these states more robust with respect to field application with increasing $l$. However, the complex phenomenology associated with these excitons has not yet fully been developed so that we will not discuss it here in further detail. 

To assess the $P$-exciton behavior more quantitatively, we have to consider the impact of the electric field in more detail. The associated potential leads to a tilting of the Coulomb potential between electron and hole along the field direction. Due to this tilting eventually tunneling of carriers through the potential can occur for sufficiently high applied voltages. Ultimately in the high field regime the exciton state is moved into the state continuum so that it can no longer be observed. This behavior determines both the absorption strength (given by the area below a resonance) and its linewidth (given by the lifetime of the state): At low fields where tunneling is still not relevant, both quantities should stay constant, while in the tunneling regime we expect an exponential dependence for both the spectrum area and linewidth. 

For determining the ionization field strength we have calculated the area below each resonance. For that purpose we have fitted each resonance with the Toyozawa formula~\cite{toyozawa,ueno,jolk} 
\begin{equation}
\label{toyozawa}
\alpha_n(E)=C_n\frac{\frac{\Gamma_n}{2}+2q_n(E-E_n)}{\left(\frac{\Gamma_n}{2}\right)^2+(E-E_n)^2}
\end{equation}
from which we obtain also the linewidth. Here $E_n$ and $\Gamma_n$ are the energy and the damping of the state $n$. $q_n$ describes the asymmetry of the observed resonance and $C_n$ gives its amplitude. Here we have restricted to the voltage regime in which the $P$-exciton line represents a well isolated line that is not influenced by adjacent absorption features. This is the regime in which the emerging $F$-excitons are still much weaker in comparison. Once higher angular momentum excitons become prominent, the spectra on the high energy flank of each $P$-exciton show pronounced modulations which we attribute to interactions with the electron-hole continuum that are not fully understood so far. This modulation also aggravates a lineshape analysis so that we refrain from a line analysis in this regime. In Fig.~\ref{Fig10-highresolutionspectrainefieldn=11} this regime corresponds to voltages higher than 0.25~V and is decreasing for higher $n$. Then the area below each absorption peak indeed shows to a good approximation an exponential drop with applied voltage as soon as effects of the applied voltage set in, up to the point where the line disappears. 

\begin{figure}[h]
\includegraphics[width=\linewidth]{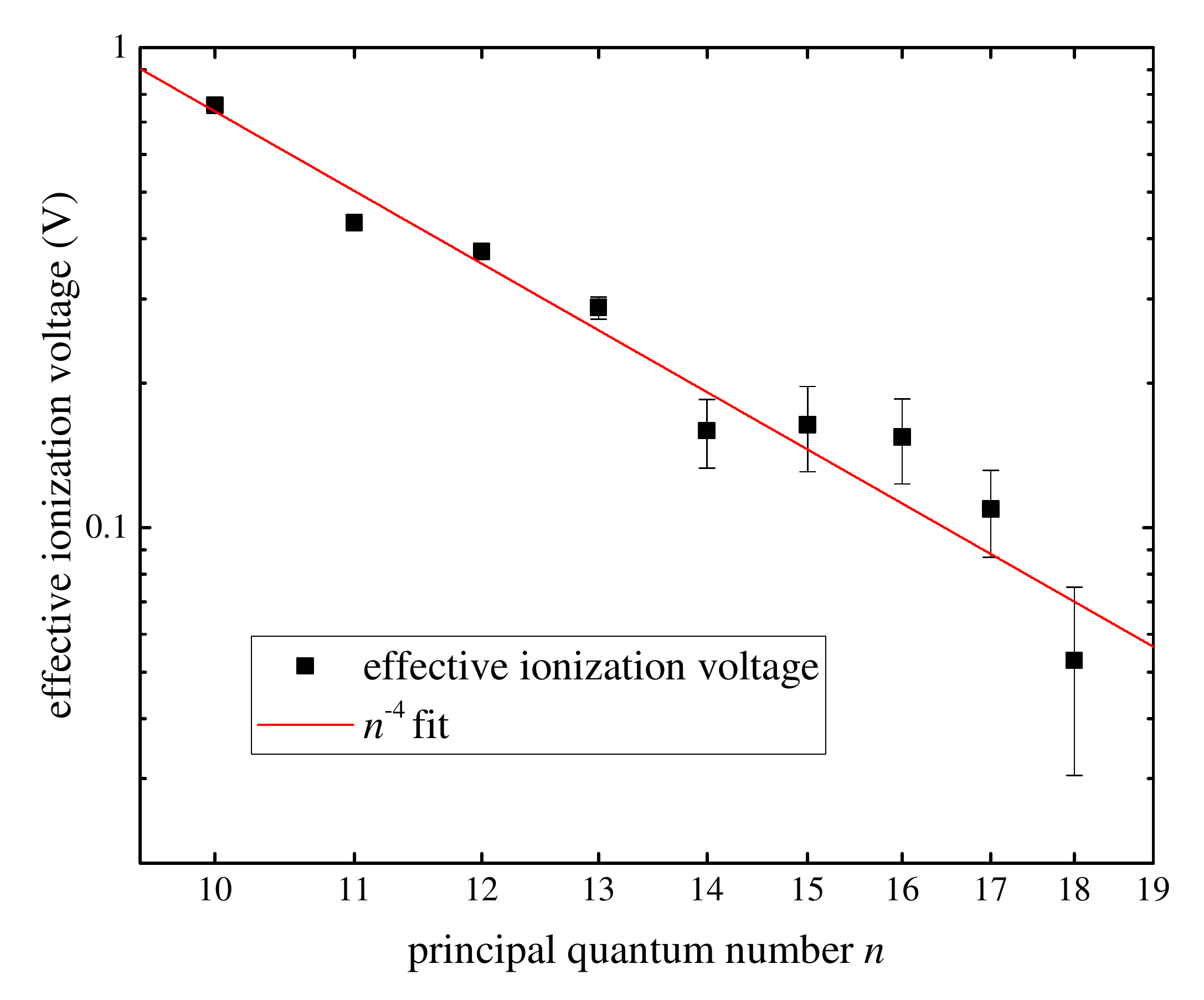}
\caption{Dependence of the dissociation voltage, determined by taking the $1/e$ value of the area below a resonance, on the principal quantum number. The solid line shows a fit to the data by a $n^{-4}$ dependence.}\label{Fig11-ionizationvoltage}
\end{figure}

From this dependence we have determined the effective ionization voltage taken as the voltage at which the area below the resonance has dropped to $1/e$. 
The drop has in fact two origins: (1) the redistribution of oscillator strength to other excitons due to state mixing; (2) the actual dissociation of the excitons. As all states are affected similarly by these two factors, we still use the voltage determined in the described way as characteristic for ionization. This voltage is plotted in Fig.~\ref{Fig11-ionizationvoltage} as function of the principal quantum number. The data can be well described by a dependence proportional to $n^{-4}$. This is the dependence that is also expected by setting Eq.~\eqref{Enl:F} for the exciton energy $E_{n,l}$ in electric field to zero. 

\begin{figure}[h]
\includegraphics[width=\linewidth]{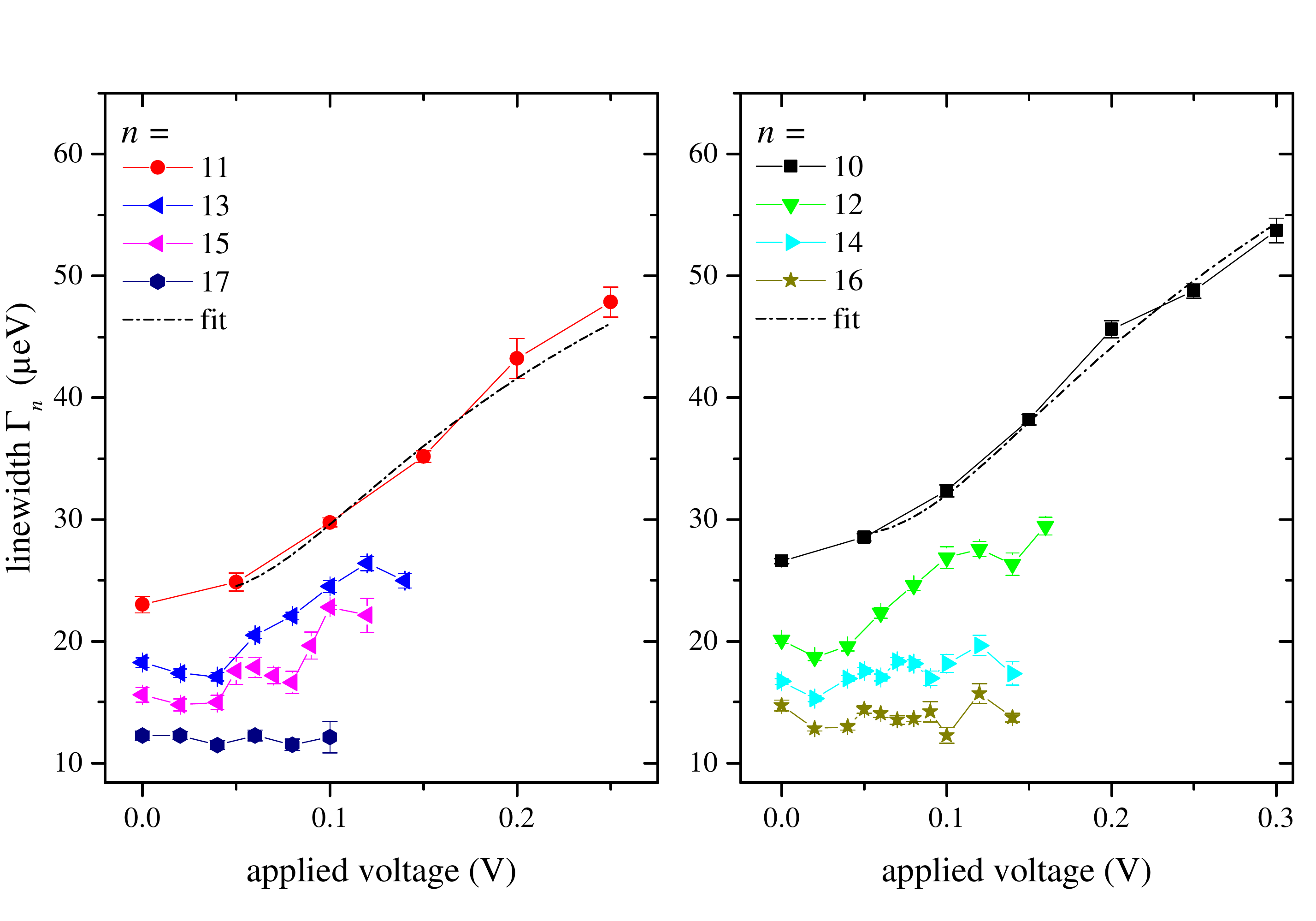}
\caption{Dependence of the linewidth of the $P$-exciton resonances on the applied voltage for different principal quantum numbers from $n=10$ up to $n=16$. The dashed lines show a fit to the data of the excitons $n=10$ and $n=11$ according to Eq.~\eqref{Gamma}. For clarity in the left (right) panel the even (odd) $n$ states are shown.}\label{Fig12-linewidthsvsvoltage}
\end{figure}

\emph{Exciton linewidth.} Finally, we turn to the dependence of the linewidth on the electric field strength, for which  one would expect -- as described above -- a transition for increasing voltage from being constant for negligible tunneling to strongly increasing as soon as tunneling can occur until ultimately the potential barrier is lowered to an extent that the excitons become unbound, which is given by the field strength at which the energy in Eq.~\eqref{Enl:F} is zero. The experimental data for the excitons from $n=10$ up to $n=16$ are shown in Fig.~\ref{Fig12-linewidthsvsvoltage}. At zero electric field we observe the well known decrease of the linewidth with increasing $n$ that has been described already in Ref.~\cite{Kazimierczuk:2014yq}.  For sufficiently low excitation powers such that effects like power broadening can be disregarded, the linewidth dependence on the principal quantum number can be described by a $n^{-3}$ law due to the corresponding scaling of the transition rates for photon and phonon emission. 

When applying the electric field, we find that the low lying excitons in the series indeed follow roughly the expected dependence (dashed line in Fig.~\ref{Fig12-linewidthsvsvoltage}) which according to the theoretical calculations should be given by~\cite{merkulov_ion,excitons:RS}
\begin{equation}
\label{Gamma}
\Gamma \propto \exp{\left(-\frac{2}{3n^3F}\right)},
\end{equation}
see Appendix~\ref{app:ion} for details. Somewhat surprisingly, the linewidth shows a different behavior for the high lying excitons in the shown set of states. Their linewidth stays constant within the experimental accuracy over the range of fields where they can be observed until they disappear.

This behavior can be understood as follows. The exponent becomes significant at the critical field
\begin{equation}
\label{Fg}
F_\Gamma \sim \frac{1}{n^3}.
\end{equation}
Since this field is parametrically larger than the ionization field $F_i$ for the states with large $n$, the increase in the linewidth is not seen, because of faster ionization of the state.

\begin{table*}[t]
\caption{\label{tab:comparison}Comparison of scaling laws with principal quantum number $n$ for Rydberg atoms and Rydberg excitons.}
\begin{ruledtabular}
\begin{tabular}{ccc}
 & Rydberg atoms & Rydberg excitons \\ \hline
\emph{Zero field} & & \\
Multiplet splitting due to quantum defect & $\propto n^{-3}$ (except of hydrogen) & $\propto n^{-3}$ \\ \hline 
\emph{Electric field} & & \\ 
Polarizability & $\propto n^{-7}$ ($\propto n^{-6}$ for hydrogen) & $\propto n^{-7}$ \\
Resonance field of states from multiplets $n$ and $n+1$&$\propto n^{-5}$&$\propto n^{-5}$\\ 
Anticrossing energy at first resonance & $\propto n^{-4}$ & $\propto n^{-4}$ \\
Ionization voltage & $\propto n^{-4}$ & $\propto n^{-4}$ \\
\hline
\emph{Magnetic field} & & \\ 
Crossover field to magnetoexciton & $\propto n^{-3}$  & $\propto n^{-3}$ \\
Resonance field of states from multiplets $n$ and $n+1$&$\propto n^{-6}$&$\propto n^{-4}$\\ 
\end{tabular}
\end{ruledtabular}
\end{table*}

\section{Conclusion}\label{sec:concl}

In summary, we have studied the scaling of several characteristic quantities with the principal quantum number in the series of Rydberg excitons in cuprous oxide
and drawn the comparison to Rydberg atoms. The parameters considered are related to the energy range covered by states at zero field due to finite quantum defects and resonant field strengths in external field. They are comprised in Table \ref{tab:comparison}. The comparison shows that for most of the considered parameters the scaling laws are identical, even though there are differences in absolute magnitude due to the strikingly different Rydberg energy. Despite of the same scaling, the origin of the parameter may be quite different like for the zero field multiplet splitting: For atoms the deviation from a $1/r$ potential leads to this splitting described by the quantum defect, while for excitons the splitting originates mostly from the complex valence band structure deviating from a quadratic dispersion law. 

When an external field is applied, identical scaling laws hold for atoms and excitons for all parameters, but in magnetic field we find differences between the two systems, mostly related to the different optical selection rules that lead to a different scaling law for the resonance field strength of the states from the multiplets with principal quantum number $n$ and $n+1$. However, even when the same scaling law is found like for the energy splitting at the resonance, the physics origin might be quite different as its size in the exciton case is determined by the crystal specific of the Hamiltonian from the hydrogen case.

From first sight the exciton level spectrum appears to be very complex, particularly under field application. Here, the demonstrated scaling laws are particularly appealing as they represent some generally valid rules for the exciton spectrum which help to develop systematics in the diversely looking spectra. The scaling laws may be transferred in a similar form also to other semiconductors.

\acknowledgements
We gratefully acknowledge the support of this project by Deutsche Forschungsgemeinschft (DFG) and the Russian Foundation of Basic Research in the frame of TRR 160(projects A1 and 	15-52-12012) and by the DFG in the SPP GiRyd. MMG is grateful to the Dynasty foundation and RF President grant MD-1555.2017.2 for partial support. MB acknowledges support by RF Government Grant No. 14.Z50.31.0021.

\appendix

\section{Landau level transitions}\label{app:ll}

\begin{figure*}[ht]
\includegraphics[width=\textwidth]{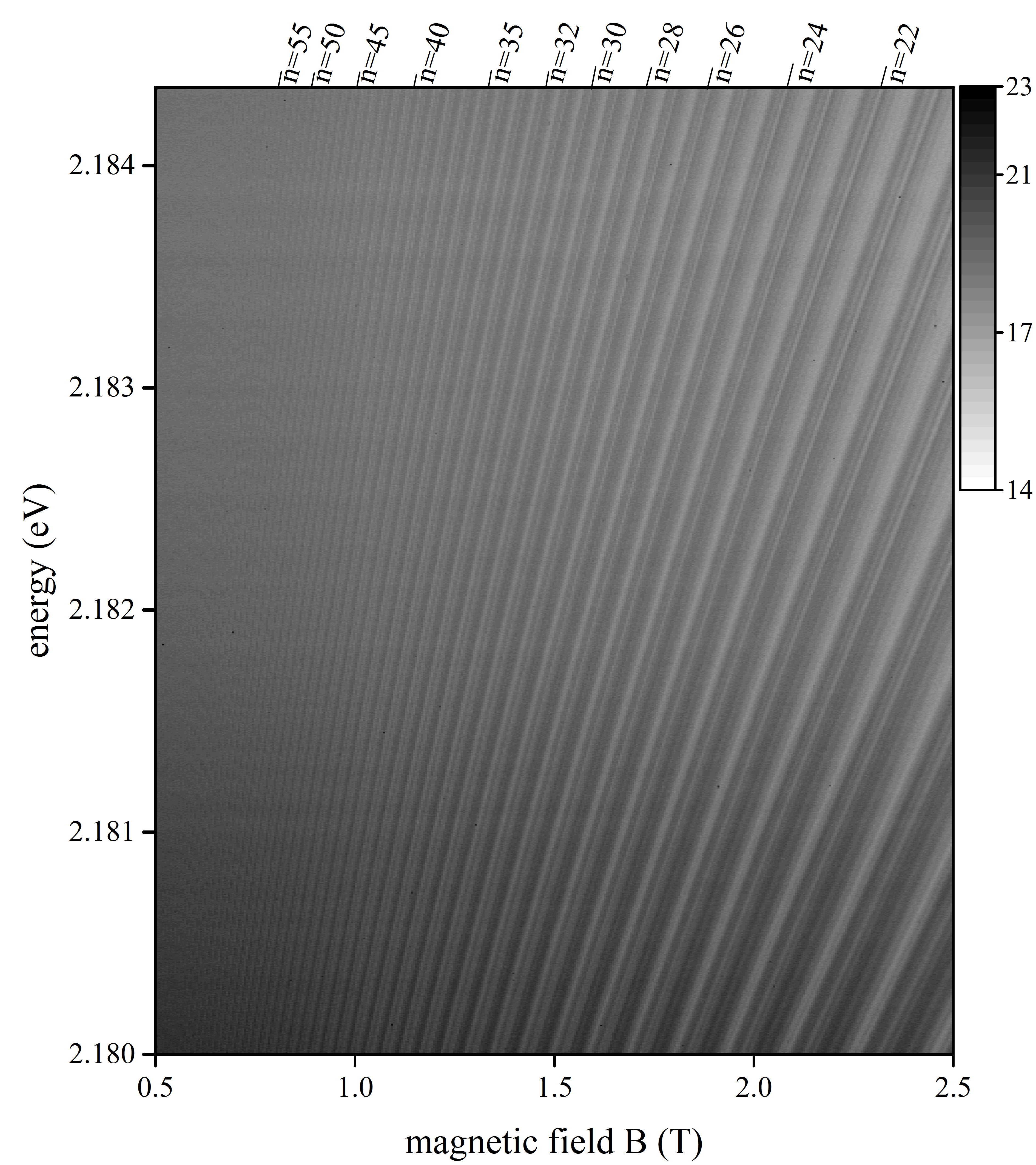}
\caption{Close-up of the black boxed region of the magnetic field spectra in Fig.~\ref{Fig1-bfield-spectra}, which allows one to resolve higher Landau level transitions. Zooming in even further allows detection of transitions associated with Landau level quantum numbers of more than 70.}\label{Fig-Appendix-LandauLevels}
\end{figure*}

Figure~\ref{Fig-Appendix-LandauLevels} zooms into the region of absorption spectra indicated by the black box in Fig.~\ref{Fig1-bfield-spectra}, from which the resonances that can be in lowest approximation assigned to transitions between electron and hole Landau levels with high quantum number can be resolved in more detail. This assignment is only approximate as for the corresponding energy range the system shows quantum chaotic behavior with multiple anticrossings, so that no strict $\bm B$-linear behavior of spectral line energies with some Coulomb related modification can be identified. Still the transitions bunch around these pure Landau level transitions so that a corresponding identification becomes possible.

From the representation in Fig.\ref{Fig-Appendix-LandauLevels} we can identify transitions up to Landau level quantum number 57. Zooming in further reveals transitions up to quantum number $n$=79. Transitions associated with a particular Landau level quantum number $n$ arise with increasing magnetic field from the $B=0$ transitions associated with $P$-excitons with the same principal quantum number $n$. Experiments have shown that under the chosen experimental conditions it is very hard to observe $P$-excitons with principal quantum number higher than 
25 at $T$=1.2~K without magnetic field, not only because the energy spacing is small well below 1~meV, but also because of the decreasing oscillator strength with increasing $n$. Field application increases the splitting between the levels and enhances the oscillator strength through squeezing the exciton wave function. However, at $B$ = 0.5~T the extension of the $n$=73 Landau level orbit is still about 300~nm.

\section{Exciton ionization in electric field}\label{app:ion}

Equation~\eqref{Enl:F} allows us also to estimate the ionization field strength from the condition $E_{nl}=0$:
\begin{equation}
\label{Fi}
F_i \sim \frac{1}{n^4}, 
\end{equation}
Note that the precise numerical coefficient depends on the angular momentum of the state due to the centrifugal barrier~\cite{gallagher2005rydberg}. 

The estimates~\eqref{Fi} and \eqref{Gamma} can be obtained in the quasi-classical picture considering the trajectory corresponding to an energy below barrier. With exponential accuracy it is sufficient to study the one-dimensional motion along the field. Two turning points can be found from the equation
\[
\frac{1}{2n^2} = 1/z+\tilde Fz, 
\]
which yields
\[
 z_{1,2} = \frac{1/(2n^2)\pm \sqrt{1/(2n^2)^2-4\tilde F}}{2F}.
\]
In weak fields
\[
z_1 \approx 2n^2, \quad z_2 \approx \frac{1}{2\tilde F n^2}.
\]
The barrier vanishes if $z_1=z_2$. Hence $\tilde F=1/(4n^4)$ yielding the ``classical'' ionization condition (the numerical coefficient can be specified more accurately following Ref.~\cite{gallagher2005rydberg}). The quasi-classical motion under the barrier provides the following action ($z_1\to 0$, triangular barrier approximation):
\[
S= \int_{z_1}^{z_2} \sqrt{1/z+Fz-1/(2n^2)} dz \sim F^{1/2}z_2^{3/2} \sim \frac{1}{Fn^3},
\]
which gives the field and principal quantum number dependence of the exponent in Eq.~\eqref{Gamma}.


%

\end{document}